\documentclass[twocolumn,trackchanges]{aastex63}

\shortauthors{Sameshima et al.}

\usepackage{booktabs,amsmath}

\begin{document}

\title{\ion{Mg}{2} and \ion{Fe}{2} fluxes of luminous quasars at $z \sim
2.7$ and evaluation of the Baldwin effect in the flux-to-abundance
conversion method for quasars}

\correspondingauthor{Hiroaki Sameshima}
\email{hsameshima@ioa.s.u-tokyo.ac.jp}

\author[0000-0001-6401-723X]{Hiroaki Sameshima}
\affiliation{Institute of Astronomy, School of Science, The University
of Tokyo, 2-21-1 Osawa, Mitaka, Tokyo 181-0015, Japan}

\author{Yuzuru Yoshii}
\affiliation{Institute of Astronomy, School of Science, The University
of Tokyo, 2-21-1 Osawa, Mitaka, Tokyo 181-0015, Japan}
\affiliation{Steward Observatory, University of Arizona, 933 North
Cherry Avenue, Room N204, Tucson, AZ 85721-0065, USA}

\author{Noriyuki Matsunaga}
\affiliation{Department of Astronomy, Graduate School of Science, The
University of Tokyo, 7-3-1 Hongo, Bunkyo-ku, Tokyo 113-0033, Japan}
\affiliation{Laboratory of Infrared High-resolution spectroscopy (LiH), Koyama
Astronomical Observatory, Kyoto Sangyo University, Motoyama, Kamigamo,
Kita-ku, Kyoto 603-8555, Japan}

\author{Naoto Kobayashi}
\affiliation{Kiso Observatory, Institute of Astronomy, School of
Science, The University of Tokyo, 10762-30 Mitake, Kiso-machi, Kiso-gun,
Nagano, 397-0101, Japan}
\affiliation{Institute of Astronomy, School of Science, The University of
Tokyo, 2-21-1 Osawa, Mitaka, Tokyo 181-0015, Japan}
\affiliation{Laboratory of Infrared High-resolution spectroscopy (LiH), Koyama
Astronomical Observatory, Kyoto Sangyo University, Motoyama, Kamigamo,
Kita-ku, Kyoto 603-8555, Japan}

\author[0000-0003-2380-8582]{Yuji Ikeda}
\affiliation{Laboratory of Infrared High-resolution spectroscopy (LiH), Koyama
Astronomical Observatory, Kyoto Sangyo University, Motoyama, Kamigamo,
Kita-ku, Kyoto 603-8555, Japan}
\affiliation{Photocoding, 460-102 Iwakura-Nakamachi, Sakyo-ku, Kyoto, 606-0025, Japan}

\author{Sohei Kondo}
\affiliation{Kiso Observatory, Institute of Astronomy, School of
Science, The University of Tokyo, 10762-30 Mitake, Kiso-machi, Kiso-gun,
Nagano, 397-0101, Japan}
\affiliation{Laboratory of Infrared High-resolution spectroscopy (LiH), Koyama
Astronomical Observatory, Kyoto Sangyo University, Motoyama, Kamigamo,
Kita-ku, Kyoto 603-8555, Japan}

\author[0000-0002-6505-3395]{Satoshi Hamano}
\affiliation{National Astronomical Observatory of Japan, 2-21-1 Osawa,
Mitaka, Tokyo 181-8588, Japan}
\affiliation{Laboratory of Infrared High-resolution spectroscopy (LiH), Koyama
Astronomical Observatory, Kyoto Sangyo University, Motoyama, Kamigamo,
Kita-ku, Kyoto 603-8555, Japan}

\author{Misaki Mizumoto}
\affiliation{Hakubi Center, Kyoto University, Yoshida-honmachi,
Sakyo-ku, Kyoto 606-8501, Japan}
\affiliation{Department of Astronomy, Graduate School of Science, Kyoto
University, Kitashirakawa-oiwakecho, Sakyo-ku, Kyoto 606-8502, Japan}

\author{Akira Arai}
\affiliation{Laboratory of Infrared High-resolution spectroscopy (LiH), Koyama
Astronomical Observatory, Kyoto Sangyo University, Motoyama, Kamigamo,
Kita-ku, Kyoto 603-8555, Japan}

\author{Chikako Yasui}
\affiliation{National Astronomical Observatory of Japan, California
office, 100 W. Walnut St., Suite 300, Pasadena, CA 91124, USA}
\affiliation{Laboratory of Infrared High-resolution spectroscopy (LiH),
Koyama Astronomical Observatory, Kyoto Sangyo University, Motoyama,
Kamigamo, Kita-ku, Kyoto 603-8555, Japan}

\author{Kei Fukue}
\affiliation{Laboratory of Infrared High-resolution spectroscopy (LiH), Koyama
Astronomical Observatory, Kyoto Sangyo University, Motoyama, Kamigamo,
Kita-ku, Kyoto 603-8555, Japan}

\author[0000-0003-2011-9159]{Hideyo Kawakita}
\affiliation{Laboratory of Infrared High-resolution spectroscopy (LiH), Koyama
Astronomical Observatory, Kyoto Sangyo University, Motoyama, Kamigamo,
Kita-ku, Kyoto 603-8555, Japan}
\affiliation{Department of Astrophysics and Atmospheric Sciences, Faculty of Sciences, Kyoto Sangyo
University, Motoyama, Kamigamo, Kita-ku, Kyoto 603-8555, Japan}

\author{Shogo Otsubo}
\affiliation{Laboratory of Infrared High-resolution spectroscopy (LiH), Koyama
Astronomical Observatory, Kyoto Sangyo University, Motoyama, Kamigamo,
Kita-ku, Kyoto 603-8555, Japan}

\author{Giuseppe Bono}
\affiliation{Dipartimento di Fisica, Universit\`{a} di Roma Tor Vergata,
via della Ricerca Scientifica 1, I-00133 Roma, Italy}
\affiliation{INAF-Osservatorio Astronomico di Roma, via Frascati 33,
I-00078 Monte Porzio Catone, Italy}

\author[0000-0002-5878-5299]{Ivo Saviane}
\affiliation{European Southern Observatory, Alonso de Cordova 3107,
Santiago, Chile}





   \begin{abstract}
    To investigate the chemical abundance of broad-line region clouds in
    quasars at high redshifts, we performed near-infrared spectroscopy
    of six luminous quasars at $z \sim 2.7$ with the WINERED
    spectrograph mounted on the New Technology Telescope (NTT) at the La
    Silla Observatory, Chile.  The measured \ion{Fe}{2}/\ion{Mg}{2} flux
    ratios nearly matched with the published data for $0.7 \lesssim z
    \lesssim 1.6$, suggesting that there is no evolution over a long
    period of cosmic time, which is consistent with previous studies.
    To derive the chemical abundances from the measured equivalent
    widths (EWs), their dependence on nonabundance factors must be
    corrected.  In our previous paper, we proposed a method to derive
    the [Mg/Fe] abundance ratio and the [Fe/H] abundance by correcting
    the dependence of EW(\ion{Mg}{2}) and EW(\ion{Fe}{2}) on the
    Eddington ratio.  To the best of our knowledge, that was the first
    report to discuss the star-formation history through a direct
    comparison with chemical evolution models.  In the present study, we
    further investigated the dependence of EWs on luminosity, which is
    known as the Baldwin effect (BEff).  Additional correction for the
    BEff significantly affects the derived chemical abundances for the
    six luminous quasars at $z \sim 2.7$, and the resultant abundances
    agree well with the prediction of chemical evolution models.  Given
    that most distant quasars found thus far are biased toward luminous
    ones, the correction of the measured EWs for the BEff is crucial for
    extending the chemical evolution study to higher redshifts.

   \end{abstract}

\keywords{galaxies: abundances --- galaxies: active --- quasars:
emission lines --- galaxies: evolution --- cosmology: observations ---
early universe}


\section{Introduction}

Many researchers have investigated the chemical evolution of the
universe using distant quasars at high redshifts.  This is mostly
because the associated history of formation and evolution of massive
stars is expected to drive the chemical evolution, as well as circulate
gaseous material and provide thermal and kinetic energy to the early
universe (e.g.,
\citealt{2014MNRAS.444.3684V,2015MNRAS.446..521S,2018MNRAS.475..648P}).

Heavy elements in the broad-line region (BLR) clouds of quasars are
produced by the supernova explosion of stars that are formed in host
galaxies (\citealt{1999ARA&A..37..487H}).  $\alpha$ elements such as Mg
are ejected from Type~II supernovae (SNe~II), which are formed when
short-lived massive stars die.  In contrast, a considerable amount of Fe
originates from Type~Ia supernovae (SNe~Ia), which are formed when
long-lived intermediate-mass stars in a binary system explode.  Thus, Fe
enrichment is significantly delayed compared with $\alpha$ elements.
This causes a break in the [$\alpha$/Fe]\footnote{$[\alpha/\mathrm{Fe}]
\equiv \log(n_\alpha/n_\mathrm{Fe}) -
\log(n_\alpha/n_\mathrm{Fe})_{\odot}$, where $n_x$ represents the number
density of the element $x$ ($=\alpha$ element or Fe).  Throughout this
paper, we will use this logarithmical expression for abundance ratios.}
ratio at the cosmic time elapsed since the formation of the first stars,
$t-t_0$, which corresponds to the typical lifetime $t_\mathrm{Ia}$ of an
SN~Ia progenitor.  This [$\alpha$/Fe] break as a nucleosynthetic
conjecture (\citealt{1979ApJ...229.1046T}) has been confirmed by
observations of metal-poor stars in the Galactic thin and thick disks
(e.g., \citealt{1997ARA&A..35..503M,2003A&A...397L...1F}) and in dwarf
spheroidal galaxies (e.g., \citealt{2009ARA&A..47..371T}).  However, a
clear signature of the [$\alpha$/Fe] break has not been observed yet in
distant quasars, and it is being actively explored for high redshifts
(e.g.,
\citealt{2014ApJ...790..145D,2017ApJ...849...91M,2020ApJ...898..105O}).

Thus far, measuring the emission-line flux ratio has been considered a
promising method for deriving [$\alpha$/Fe]; the ultraviolet (UV)
emission lines of \ion{Mg}{2} $\lambda2798$ and \ion{Fe}{2}, which
exhibits a broad composite feature at $\sim$2000--3000~\AA, have been
considered as an ideal line pair (e.g., \citealt{1985ApJ...288...94W}).
Thus the \ion{Fe}{2}/\ion{Mg}{2} flux ratio of quasars has been measured
over a wide range of redshift extending to $z \sim 7$ by many
researchers (e.g.,
\citealt{1996ApJ...470L..85K,1999ApJ...515..487T,2003ApJ...587L..67F,2003ApJ...594L..95B,2003ApJ...596L.155M,2002ApJ...564..581D,2003ApJ...596..817D,2002ApJ...565...63I,2004ApJ...614...69I,2006ApJ...650...57T,2007AJ....134.1150J,2007ApJ...669...32K,2009MNRAS.395.1087S,2011ApJ...739...56D,2014ApJ...790..145D,2017ApJ...849...91M,2019ApJ...874...22S,2020ApJ...898..105O}).
However, the measured flux ratios deviate significantly beyond the
measurement error range,\footnote{A part of the variance is arguably due
to the difference in the fitting procedures adopted.  Therefore, it is
advisable to obtain spectra of similar quality and analyze them in the
same manner.} preventing us from finding any signature of the [Mg/Fe]
evolution over a range of redshift explored.  Hence, it was uncertain
whether the assumption that the \ion{Fe}{2}/\ion{Mg}{2} flux ratio was a
first-order proxy for [Mg/Fe] was correct (e.g.,
\citealt{2011ApJ...736...86D,2011ApJ...739...56D}).

As a turning point in this research area, we developed a method for
abundance diagnostics of BLR clouds (\citealt{2017ApJ...834..203S};
hereafter, Paper I), which compares the measured equivalent widths (EWs)
of \ion{Mg}{2} and \ion{Fe}{2} that are corrected for their dependence
on the Eddington ratio with detailed photoionization simulations.  It
was shown, for the first time, that the derived [Mg/Fe] abundance ratio
and [Fe/H] abundance agreed well with the prediction of chemical
evolution models for the redshift range of $0.7 \lesssim z \lesssim
1.6$.  However, because the chemical evolution models converge at $z
\lesssim 2$ regardless of the star-formation history in the early
universe, the most plausible model has not been determined yet.

Thus, high-redshift data are vital to overcome the indistinguishability
at $z \lesssim 2$ in chemical evolution models.  High-redshift quasars
that have been found so far are biased toward luminous ones owing to
observational limitations.  It is widely known that for quasars, an
anticorrelation was observed between an emission-line EW and luminosity,
which is called the Baldwin effect (BEff;
\citealt{1977ApJ...214..679B}).  First, we report the results of
near-infrared spectroscopic observations of six luminous quasars at $z
\sim 2.7$, and later we discuss how the influence of the BEff on our
abundance diagnostics can be excluded.  This will enable us to derive
the chemical abundance of BLR clouds at a much higher redshift than that
studied in Paper I.

The remainder of this paper is organized as follows.  Our target
selection and the details of observations are presented in Section~2.
Data reduction including one-dimensional spectrum extraction and
correction for telluric absorption is described in Section~3.  The
measurement of emission lines by fitting a spectral model to the
observed spectrum is described in Section~4, and the results are
summarized in Section~5.  In Section~6, the investigation of the BEff
for \ion{Mg}{2} and \ion{Fe}{2} emission lines using a large sample of
Sloan Digital Sky Survey (SDSS) quasars is presented, and its
relationship with the dependence on the Eddington ratio is discussed.
Further, a method to correct the BEff to derive [Mg/Fe] and [Fe/H] is
presented, and it is compared with the chemical evolution models in the
redshift range of $0.7 < z < 2.7$.  The paper is summarized in
Section~7.  Throughout this paper, we assume $\Lambda$CDM cosmology,
with $\Omega_\Lambda = 0.7$, $\Omega_M = 0.3$, and
$H_0=70~\mathrm{km~s^{-1}~Mpc^{-1}}$.


\section{Target selection and observation} \label{sec:target}

Target candidates were selected from the SDSS Data Release 12 (DR12)
Quasar Catalog (\citealt{2017AA...597A..79P}).  To avoid serious
telluric absorption, the redshift range was restricted to approximately
2.7 so that the \ion{Mg}{2} $\lambda2798$ emission line would be located
in the central region of the $Y$ band (9760--11100\,\AA).

The observations were carried out in 2018 March using the WINERED
echelle spectrograph (\citealt{2016SPIE.9908E..5ZI}) mounted on the ESO
3.58 m New Technology Telescope (NTT) at the La Silla Observatory, Chile
(ESO program ID: 0100.B-0939(A)).  The observation mode was set to the
WIDE mode with a 200 \micron\ width slit, which had a wavelength
coverage of 0.90--1.35~\micron\ and a resolving power of $R \sim 17000$.
The targets were observed at two positions separated by approximately
10\arcsec\ along the slit by nodding the telescope to make the ABBA
dithering sequence.  Telluric standard stars were also observed before
or after the targets so that the differences in time and airmass between
them were as close as possible.  Each night, HD 111844, an F-type star
whose medium-resolution ($R \sim 2000$) spectrum was available at the
IRTF Spectral
Library\footnote{\url{http://irtfweb.ifa.hawaii.edu/~spex/IRTF_Spectral_Library/}}
(\citealt{2009ApJS..185..289R}), was observed as the flux standard star.
Finally, we observed six quasars (hereafter, NTT quasars), some of which
were observed over multiple days.  The observation log is summarized in
Table \ref{tab:obslog}.

\begin{figure}[t]
 \epsscale{1.1}
 \plotone{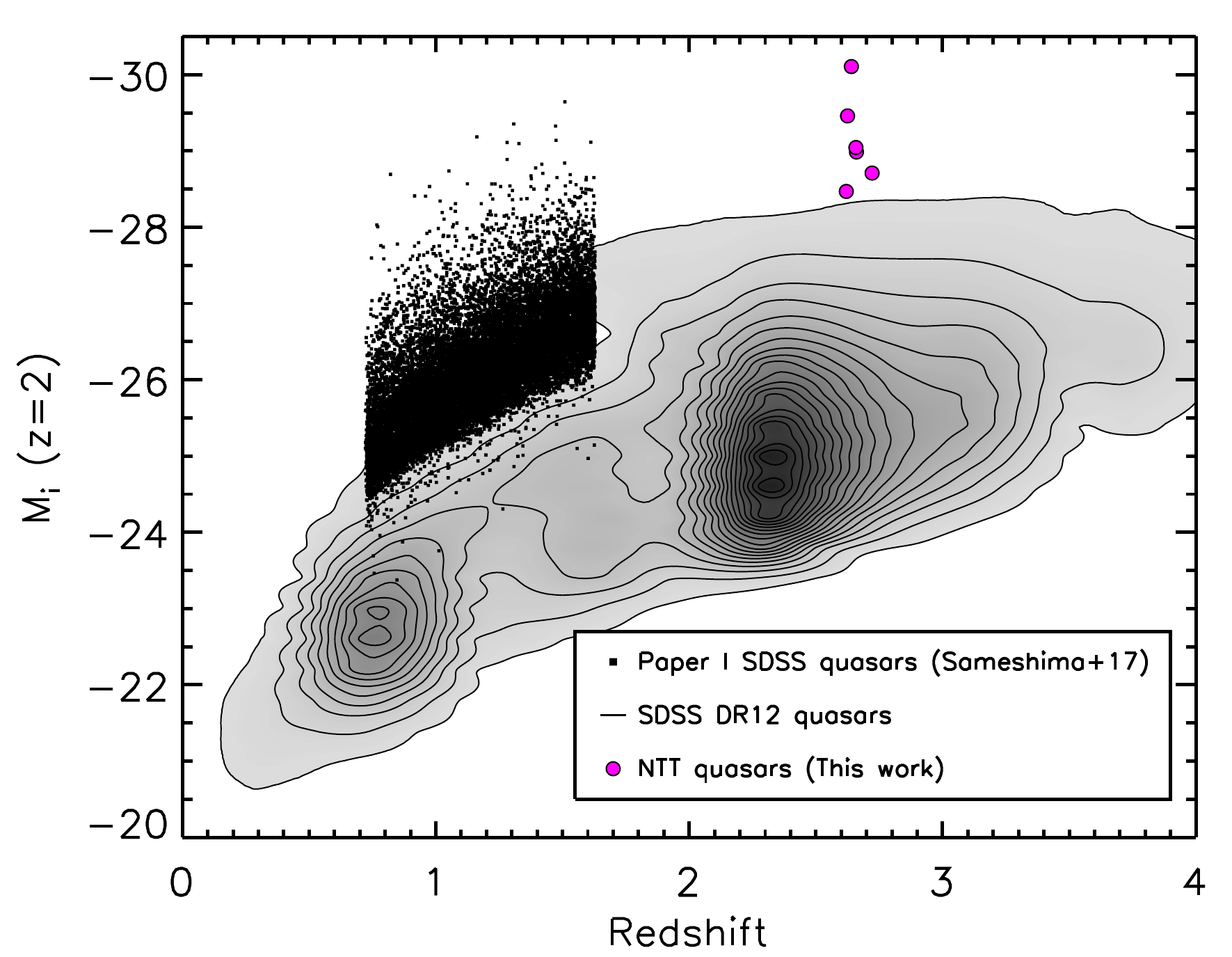}
 \caption{Sample distribution in the absolute magnitude--redshift plane
 for NTT quasars (magenta circle) and Paper~I SDSS quasars (black dot).
 For reference, the distribution of all samples in the SDSS DR12 Quasar
 Catalog is shown by contours.}
 \label{fig:mag_abs_distri}
 \end{figure}

Figure \ref{fig:mag_abs_distri} shows the sample distribution of the NTT
quasars in the absolute magnitude--redshift plane; the quasars included
in the SDSS DR12 Quasar Catalog and those investigated in Paper I
(hereafter, Paper I SDSS quasars) are also shown as reference.  From
this figure, it can be seen that the NTT quasars are much brighter than
most Paper I SDSS quasars, and they are even located in the bright end
of quasars with nearly the same redshift.  This characteristic is
suitable for investigating the BEff.

\begin{deluxetable*}{lllllllclrlr}[t]
\tabletypesize{\footnotesize}
\tablecaption{Summary of observation \label{tab:obslog}}
\tablehead{
\colhead{Object} & \colhead{$\alpha$\,(J2000.0)} & \colhead{$\delta$\,(J2000.0)} &
 \colhead{$z$\tablenotemark{a}} & \colhead{$J_\mathrm{Vega}$\tablenotemark{b}} & \multicolumn{5}{c}{WINERED} &
 \multicolumn{2}{c}{SDSS} \\
 \cmidrule(r){6-10} \cmidrule(r){11-12}
 & & & & & \colhead{Exposure} & \colhead{Airmass} & \colhead{Seeing\tablenotemark{c}} & \colhead{Date} &
 \colhead{S/N\tablenotemark{d}} & \colhead{Date\tablenotemark{e}} &
 \colhead{S/N\tablenotemark{e}}
 }
 \startdata
 J0828+1251 & 08 28 57.22 & +12 51 40.0 & 2.722 & 15.974 & 600~s $\times$ 10 & 1.39--1.38 & 1.2 & 2018 Mar 4 & 7.1     & 2011 Jan 3  & 21.53 \\
            &             &             &       &        & 600~s $\times$ 6  & 1.39--1.53 & 0.9 & 2018 Mar 5 &         &             & \\
 J0850+1108 & 08 50 45.73 & +11 08 40.5 & 2.620 & 16.470 & 900~s $\times$ 2  & 1.33--1.32 & 1.8 & 2018 Mar 2 & 13.5    & 2012 Jan 21 & 22.31 \\
            &             &             &       &        & 900~s $\times$ 8  & 1.70--1.31 & 1.1 & 2018 Mar 3 &         &             & \\
            &             &             &       &        & 900~s $\times$ 8  & 1.55--1.31 & 1.0 & 2018 Mar 5 &         &             & \\
            &             &             &       &        & 900~s $\times$ 8  & 1.33--1.41 & 0.9 & 2018 Mar 6 &         &             & \\
 J1011+2941 & 10 11 55.59 & +29 41 41.5 & 2.640 & 14.870 & 900~s $\times$ 3  & 1.98--1.93 & 0.8 & 2018 Mar 3 & 12.6    & 2013 Feb 5  & 46.99 \\
 J1107+0436 & 11 07 08.41 & +04 36 17.9 & 2.660 & 15.685 & 900~s $\times$ 10 & 1.28--1.28 & 1.0 & 2018 Mar 4 & 22.1    & 2012 Jan 2  & 39.47 \\
 J1130+0732 & 11 30 17.37 & +07 32 12.9 & 2.658 & 16.110 & 900~s $\times$ 12 & 1.28--1.53 & 0.8 & 2018 Mar 6 & 15.2    & 2012 Jan 21 & 48.24 \\
 J1142+2654 & 11 42 54.26 & +26 54 57.5 & 2.625 & 15.823 & 750~s $\times$ 8  & 1.94--1.80 & 1.0 & 2018 Mar 2 & 16.2    & 2013 Feb 13 & 46.36 \\
            &             &             &       &        & 900~s $\times$ 4  & 1.86--1.79 & 1.3 & 2018 Mar 3 &         &             & \\
 \enddata
 \tablenotetext{a}{Redshift from visual inspection (Z\_VI) taken from
 the SDSS DR12 Quasar catalog (\citealt{2017AA...597A..79P}).}
 \tablenotetext{b}{$J$-band magnitude (Vega) taken from the 2MASS All
 Sky Catalog of point sources (\citealt{2003tmc..book.....C}).}
 \tablenotetext{c}{Average seeing during the observation in the unit of arcsec.}
 \tablenotetext{d}{Median signal-to-noise ratio per pixel at the $J$ band of the
 spectrum after coadding and rebinning.}
 \tablenotetext{e}{Taken from the Sloan Digital Sky Survey III Science
 Archive Server (\url{https://dr12.sdss.org/}).}
 \end{deluxetable*}


\section{Data reduction} \label{sec:reduction}

The observed data of the NTT quasars were reduced in a standard manner
using IRAF\footnote{ IRAF is distributed by the National Optical
Astronomy Observatories, which are operated by the Association of
Universities for Research in Astronomy, Inc., under cooperative
agreement with the National Science Foundation.}  routines as follows.
First, sky subtraction was performed by taking the difference between
two consecutive images taken at different slit positions, i.e., A$-$B
and B$-$A.  Then, scattered light was evaluated in the interorder
regions of each difference image and was removed.  Flat fielding was
performed using a dome-flat image.  The bad pixels were masked and
replaced through linear interpolation from the surrounding pixels.  The
spectral line tilts in the two-dimensional images were corrected by
performing a geometrical transformation using arc-lamp images as
reference.  Then, one-dimensional spectra were extracted using the IRAF
task {\tt apall} for each order, and background subtraction was
performed by using the sky region.  Wavelength calibration was performed
using the Th--Ar lamp spectra extracted in the same manner as the target
object.  Then, each frame spectrum observed on the same day was coadded.

\begin{figure}[t]
 \epsscale{1.1}
 \plotone{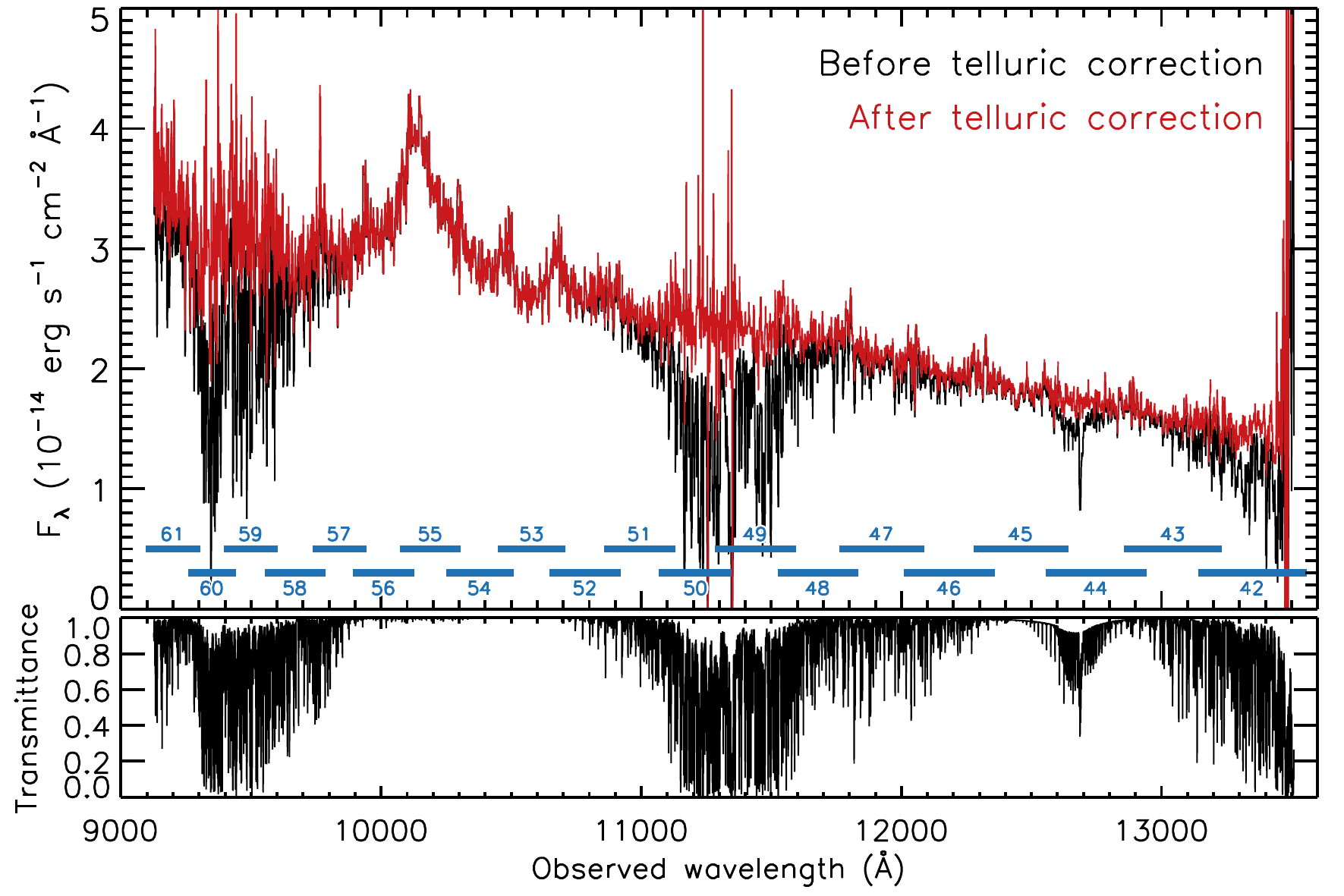}
 \caption{An example of correction for telluric absorption.  In the
 upper panel, the spectra prior to (black) and after (red) telluric
 correction are shown with the free spectral ranges of echelle orders
 (blue).  In the lower panel, a synthetic telluric spectrum created with
 {\tt molecfit} is shown as reference.}
 \label{fig:telluric_correction}
\end{figure}

The telluric absorption was corrected by using synthetic telluric
spectra created by {\tt molecfit}
(\citealt{2015A&A...576A..77S,2015A&A...576A..78K}) as follows.  Because
the signal-to-noise ratio (S/N) of the obtained quasar spectrum was not
sufficient for the accurate estimation of the telluric absorption, we
used the high-S/N spectrum of the corresponding telluric standard star
instead.  By running {\tt molecfit} with the spectrum of the telluric
standard star, we created a synthetic telluric spectrum and used it with
the IRAF task {\tt telluric} to correct the quasar spectrum for telluric
absorption.  An example of the correction for telluric absorption is
illustrated in Figure \ref{fig:telluric_correction}.

\begin{figure*}[t]
 \epsscale{1.1}
 \plotone{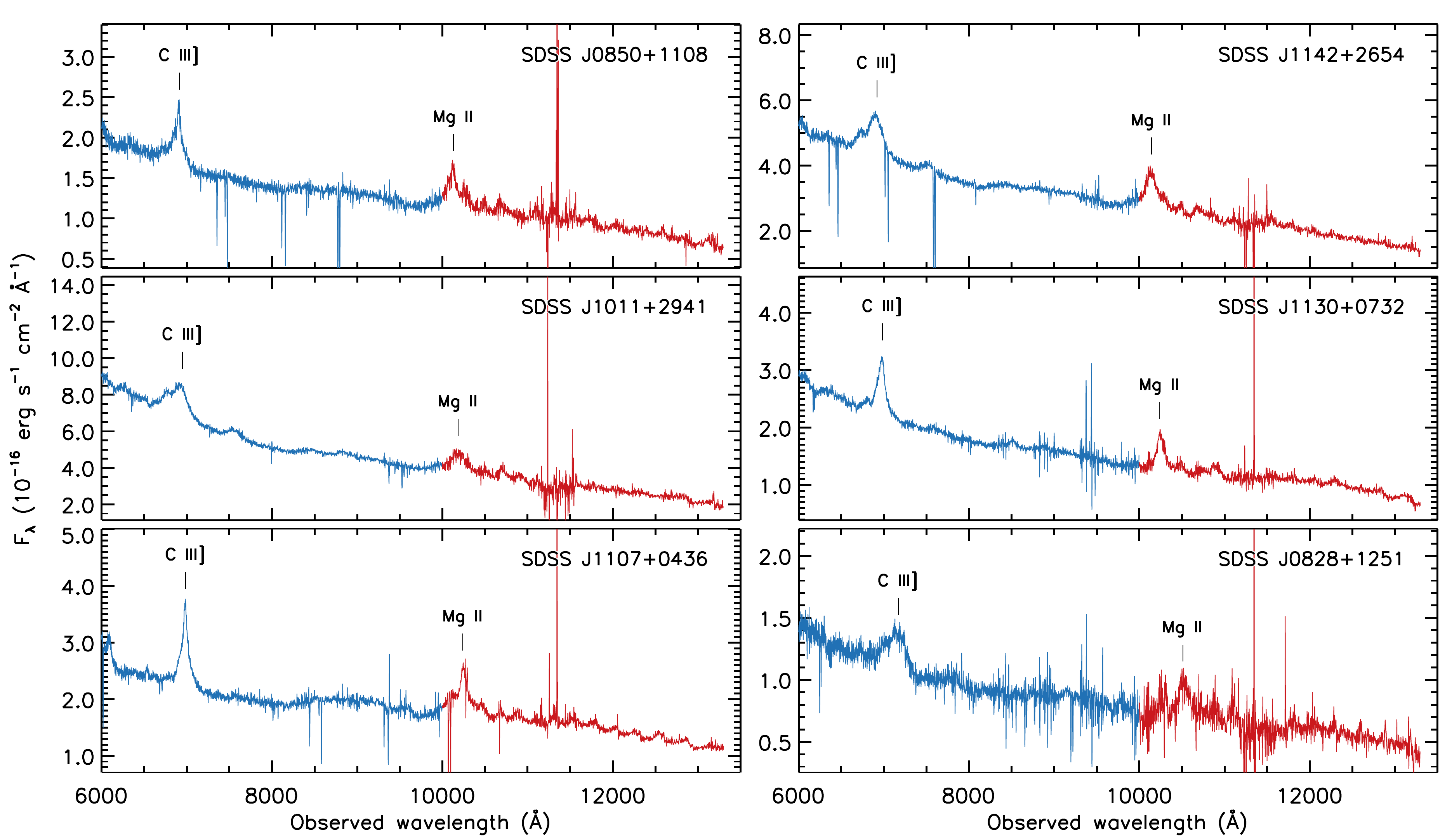}
 \caption{Observed spectra of NTT quasars.  The blue line represents the
 archival SDSS spectrum, and the red line represents the WINERED
 spectrum that is rebinned and scaled to match the SDSS spectrum.}
 \label{fig:spec}
\end{figure*}

The response curves were derived from the observed spectrum of the flux
standard star HD 111844 that was corrected for telluric absorption in
the same manner as quasars.  These response curves were used for flux
calibration to restore the true slopes of the quasar spectra.  For
quasars that were observed for multiple days, each flux-calibrated
spectrum was scaled so that the median fluxes matched, and then they
were coadded with weights according to the exposure time.  Finally, the
spectrum was rebinned by calculating the median of neighboring 13 or 14
pixels and scaled to match the corresponding low-resolution optical
spectrum taken by SDSS at $\sim$1\,\micron, assuming that the change of
spectral shape during the time between the SDSS and WINERED observations
was negligible.  All the final spectra of the NTT quasars are shown in
Figure \ref{fig:spec}, and their median S/Ns in the $J$ band are listed
in Table \ref{tab:obslog}.

\section{Analysis} \label{sec:analysis}

\begin{figure*}[t]
 \epsscale{1.1}
 \plotone{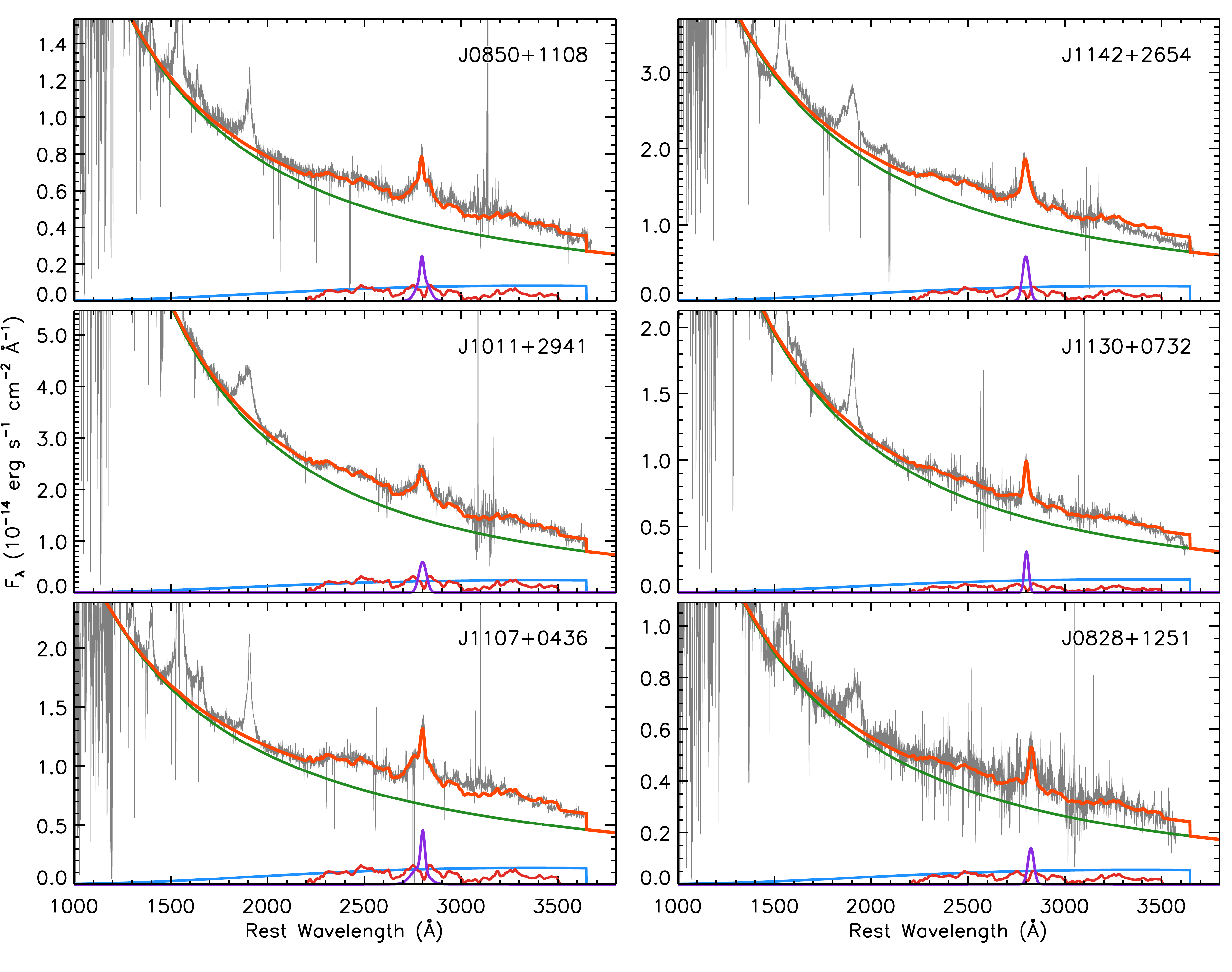}
 \caption{Results of spectral fitting for NTT quasars.  The best-fit
 power-law continuum (green), Balmer continuum (blue), \ion{Fe}{2}
 emission lines (red), \ion{Mg}{2} emission line (purple), and their
 sum (orange) are overplotted on the observed spectrum (gray).}
 \label{fig:fitting}
\end{figure*}

The fluxes of \ion{Mg}{2} and \ion{Fe}{2} emission lines were measured
in the same manner as in Paper~I.  First, the following continuum model
was fitted to the spectrum:
\begin{equation}
  F_{\lambda} = F_{\lambda}^\mathrm{PL}(\alpha,\beta) + F_{\lambda}^\mathrm{BaC} + F_{\lambda}^\mathrm{FeII}(\gamma),
\end{equation}
where $F_\lambda^{\mathrm{PL}}$ is a power-law continuum flux emitted
from an accretion disk, $F_\lambda^{\mathrm{BaC}}$ is a Balmer continuum
flux, and $F_\lambda^{\mathrm{FeII}}$ is a \ion{Fe}{2} pseudo-continuum
flux.  We adopted the Balmer continuum model proposed by
\cite{1982ApJ...255...25G} for $F_\lambda^{\mathrm{BaC}}$; the shape and
flux ratio against the power-law component were fixed as was done in
previous works (e.g.,
\citealt{2011ApJ...739...56D,2017ApJ...834..203S,2019ApJ...874...22S}).
We used the \ion{Fe}{2} template given by \cite{2006ApJ...650...57T}.
Prior to fitting, the \ion{Fe}{2} template was broadened by convolution
with a Gaussian function, for which the FWHM was fixed\footnote{Note
that, as \cite{2011ApJ...739...56D} pointed out, the FWHM adopted for
the convolved Gaussian function has a negligible effect on the measured
\ion{Fe}{2} flux.}  at 2000 km s$^{-1}$.  Thus, there were three free
parameters: the power-law slope ($\alpha$), normalization of the
power-law continuum flux ($\beta$), and normalization of the \ion{Fe}{2}
pseudo-continuum flux ($\gamma$). The best-fit parameters were obtained
by performing $\chi^2$ minimization with the IDL procedure {\tt
MPFIT.pro} (\citealt{2009ASPC..411..251M}). \ion{Fe}{2} flux was
calculated by integrating the fitted \ion{Fe}{2} template in the
wavelength range of 2200--3090\AA. Then, the \ion{Mg}{2} $\lambda2798$
emission line in the continuum-subtracted spectrum was fitted with one
or two Gaussians with the {\tt MPFIT.pro} procedure.  For fitting with
two Gaussians, both the flux and the FWHM of \ion{Mg}{2} were calculated
from the sum of the Gaussians.  The rest-frame EWs of \ion{Fe}{2} and
\ion{Mg}{2} emission lines were measured by dividing the line flux by
the continuum flux density at 3000~\AA.  Examples of spectral fitting
for the NTT quasars are shown in Figure \ref{fig:fitting}.

The mass of a black hole (BH), $M_{\mathrm{BH}}$, was estimated from
the \ion{Mg}{2} FWHM and the continuum luminosity at 3000~\AA\ by using
the virial mass estimate formula (\citealt{2009ApJ...699..800V}):
\begin{eqnarray}
\log \left( \frac{M_{\mathrm{BH}}}{M_{\odot}} \right) = 6.86 + 2 \log
 \left( \frac{\mathrm{FWHM(Mg\,II)}}{1000\ \mathrm{km\ s^{-1}}} \right)
 \nonumber \\
 + 0.5 \log \left( \frac{\lambda L_\lambda(3000\mathrm{\AA})}{10^{44}\
	     \mathrm{erg\ s^{-1}}} \right). \label{eq:bhmass}
\end{eqnarray}
For a BLR cloud orbiting the central BH, the Eddington luminosity is
given by
\begin{equation}
 L_{\mathrm{Edd}} = \frac{4\pi Gc m_p}{\sigma_e}M_{\mathrm{BH}}, \label{eq:edd_lumi}
\end{equation}
where $G$ is the gravitational constant, $c$ is the speed of light,
$m_p$ is the proton mass, and $\sigma_e$ is the Thomson scattering cross
section (\citealt{1997iagn.book.....P}).  Following
\cite{2011ApJS..194...45S}, we estimated the bolometric luminosity from
the measured monochromatic luminosity at 3000~\AA\ with the bolometric
correction formula:
\begin{equation}
 L \equiv L_{\mathrm{bol}} = 5.15 \lambda L_\lambda(3000\mathrm{\AA}). \label{eq:bol_lumi}
\end{equation}
From equations (\ref{eq:bhmass})--(\ref{eq:bol_lumi}), the Eddington ratio
is written as
\begin{eqnarray}
 \log \left( \frac{L}{L_{\mathrm{Edd}}} \right) = 
  -0.249 - 2 \log \left( \frac{\mathrm{FWHM(Mg\,II)}}{1000\ \mathrm{km\
		s^{-1}}} \right) \nonumber \\
 + 0.5 \log \left( \frac{\lambda L_\lambda(3000\mathrm{\AA})}{10^{44}\
	 \mathrm{erg\ s^{-1}}} \right). \label{eq:EddRatio}
\end{eqnarray}
For all the NTT quasars, the Eddington ratio was evaluated using the
above equation, as was done for the Paper~I SDSS quasars.

The measurement errors were estimated through Monte Carlo simulations,
similarly to the study of \cite{2019ApJ...874...22S}.  We performed the
above fitting procedures for a set of 1000 mock spectra generated by
randomizing the errors in each pixel of the observed spectrum.  For each
measured quantity, the median value of the mock spectra was adopted as
the maximum likelihood estimate, and the interval between the 15.87th
and 84.13th percentiles was adopted as the $1\sigma$ confidence
interval.  All measurements and errors of the NTT quasars calculated in
this manner are summarized in Table \ref{tab:measure}.

\begin{deluxetable*}{lrrlrrrrr}[t]
\tablecaption{Results of fitting \label{tab:measure}}
\tablehead{
 \colhead{Object} & \colhead{$z_\mathrm{Mg\,II}$\tablenotemark{a}} & \colhead{FWHM} & 
 \colhead{EW(\ion{Mg}{2})\tablenotemark{b}} & \colhead{EW(\ion{Fe}{2})\tablenotemark{b}} &
 \colhead{\ion{Fe}{2}/\ion{Mg}{2}} & \colhead{$\log L_{3000}$\tablenotemark{c}} & \colhead{$\log M_\mathrm{BH}/M_\odot$} &
 \colhead{$\log L/L_\mathrm{Edd}$} \\
 & & \colhead{(km~s$^{-1}$)} & \colhead{(\AA)} & \colhead{(\AA)} & & \colhead{(erg~s$^{-1}$)} & & 
 }
 \startdata
 J0828+1251 & $2.758^{+0.003}_{-0.002}$ & $4400^{+460}_{-400}$ & $18.1^{+1.5}_{-1.5}$  & $75.9^{+6.5}_{-6.1}$  & $4.21^{+0.56}_{-0.50}$ & $46.638^{+0.003}_{-0.003}$ & $9.47^{+0.09}_{-0.08}$ & $-0.13^{+0.17}_{-0.18}$ \\
 J0850+1108 & $2.617^{+0.002}_{-0.002}$ & $4080^{+490}_{-540}$ & $29.7^{+1.3}_{-1.3}$  & $99.1^{+3.7}_{-3.4}$  & $3.33^{+0.21}_{-0.19}$ & $46.758^{+0.002}_{-0.002}$ & $9.46^{+0.10}_{-0.12}$ & $-0.09^{+0.12}_{-0.10}$ \\
 J1011+2941 & $2.644^{+0.002}_{-0.002}$ & $5040^{+280}_{-280}$ & $19.9^{+1.0}_{-1.0}$  & $111.4^{+2.8}_{-3.0}$ & $5.60^{+0.32}_{-0.31}$ & $47.272^{+0.002}_{-0.002}$ & $9.90^{+0.05}_{-0.05}$ & $-0.02^{+0.05}_{-0.05}$ \\
 J1107+0436 & $2.666^{+0.001}_{-0.001}$ & $3000^{+180}_{-170}$ & $37.1^{+10.6}_{-4.4}$ & $111.9^{+2.9}_{-3.4}$ & $2.97^{+0.48}_{-0.68}$ & $46.981^{+0.002}_{-0.003}$ & $9.30^{+0.05}_{-0.05}$ & $+0.29^{+0.05}_{-0.05}$ \\
 J1130+0732 & $2.664^{+0.001}_{-0.001}$ & $2780^{+110}_{-110}$ & $15.4^{+0.6}_{-0.7}$  & $72.4^{+2.8}_{-2.7}$  & $4.71^{+0.29}_{-0.25}$ & $46.875^{+0.001}_{-0.002}$ & $9.18^{+0.03}_{-0.04}$ & $+0.30^{+0.03}_{-0.03}$ \\
 J1142+2654 & $2.625^{+0.001}_{-0.001}$ & $4650^{+170}_{-150}$ & $23.7^{+0.7}_{-0.7}$  & $84.7^{+1.9}_{-1.9}$  & $3.59^{+0.13}_{-0.14}$ & $47.144^{+0.001}_{-0.001}$ & $9.77^{+0.03}_{-0.03}$ & $-0.01^{+0.03}_{-0.03}$ \\
 \enddata
 \tablenotetext{a}{Redshift obtained from the fitting of the \ion{Mg}{2} emission line.}
 \tablenotetext{b}{EWs are measured in the rest-frame wavelength.}
 \tablenotetext{c}{$L_{3000} \equiv \lambda L_\lambda$(3000\AA).}
\end{deluxetable*}


\section{Results} \label{sec:result}

\subsection{Luminosity and the Eddington ratio}

\begin{figure}[t]
 \epsscale{1.1}
 \plotone{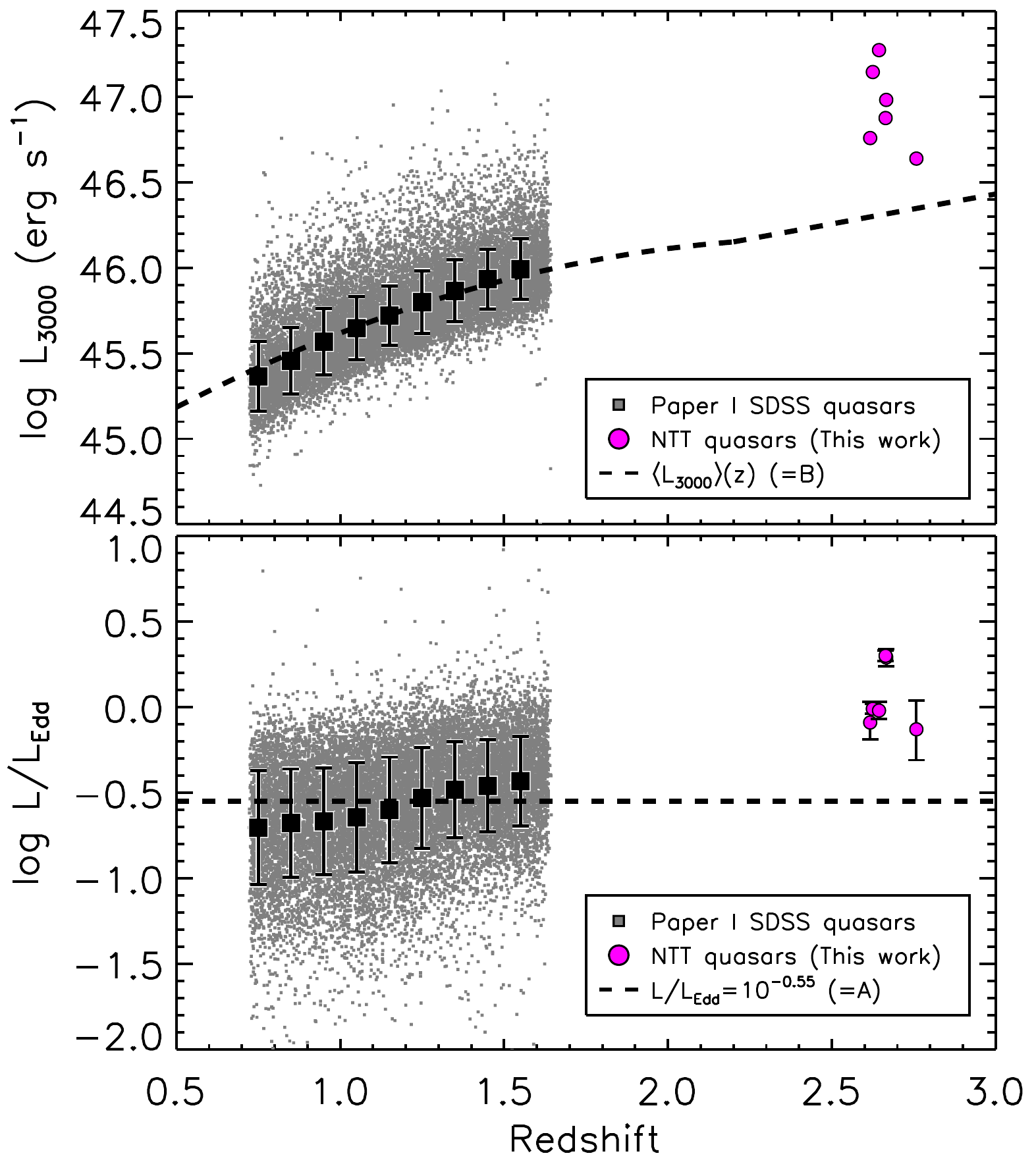}
 \caption{Distribution of the six NTT quasars at $z \sim 2.7$ (magenta
 circle) and the Paper~I data (gray dots) on the $\log L_{3000}$--$z$
 plane (top panel) and $\log L/L_\mathrm{Edd}$--$z$ plane (bottom
 panel).  The dashed lines represent the fiducial values $B$ and $A$,
 which are discussed in \S\ref{sec:abundance}.}
 \label{fig:luminosity_redshift}
\end{figure}

Figure \ref{fig:luminosity_redshift} shows the monochromatic luminosity
$L_{3000}$ ($\equiv \lambda L_\lambda$(3000\AA)) and the Eddington ratio
as a function of redshift for the six NTT quasars analyzed in this
study, along with the Paper~I SDSS quasars for comparison.  The
luminosity of the NTT quasars spans $10^{46.6} < L_{3000} < 10^{47.3}$
erg~s$^{-1}$, which is considerably brighter than most Paper~I SDSS
quasars.  Note that the WINERED spectrum was scaled to the corresponding
SDSS DR12 spectrum obtained for the same object, but a few years
earlier; therefore, the obtained values for the NTT quasars do not
necessarily reflect luminosities in the observation period of WINERED,
i.e., 2018 March.  The Eddington ratios of NTT quasars are systematically
greater than those of Paper~I SDSS quasars, which is mainly due to the
brightness.  Further, almost all NTT quasars fulfill $L/L_\mathrm{Edd}
\gtrsim 1$, suggesting that they are likely to be super-Eddington
objects.

\subsection{\ion{Fe}{2}/\ion{Mg}{2} fluxes}

\begin{figure}[t]
 \epsscale{1.1}
 \plotone{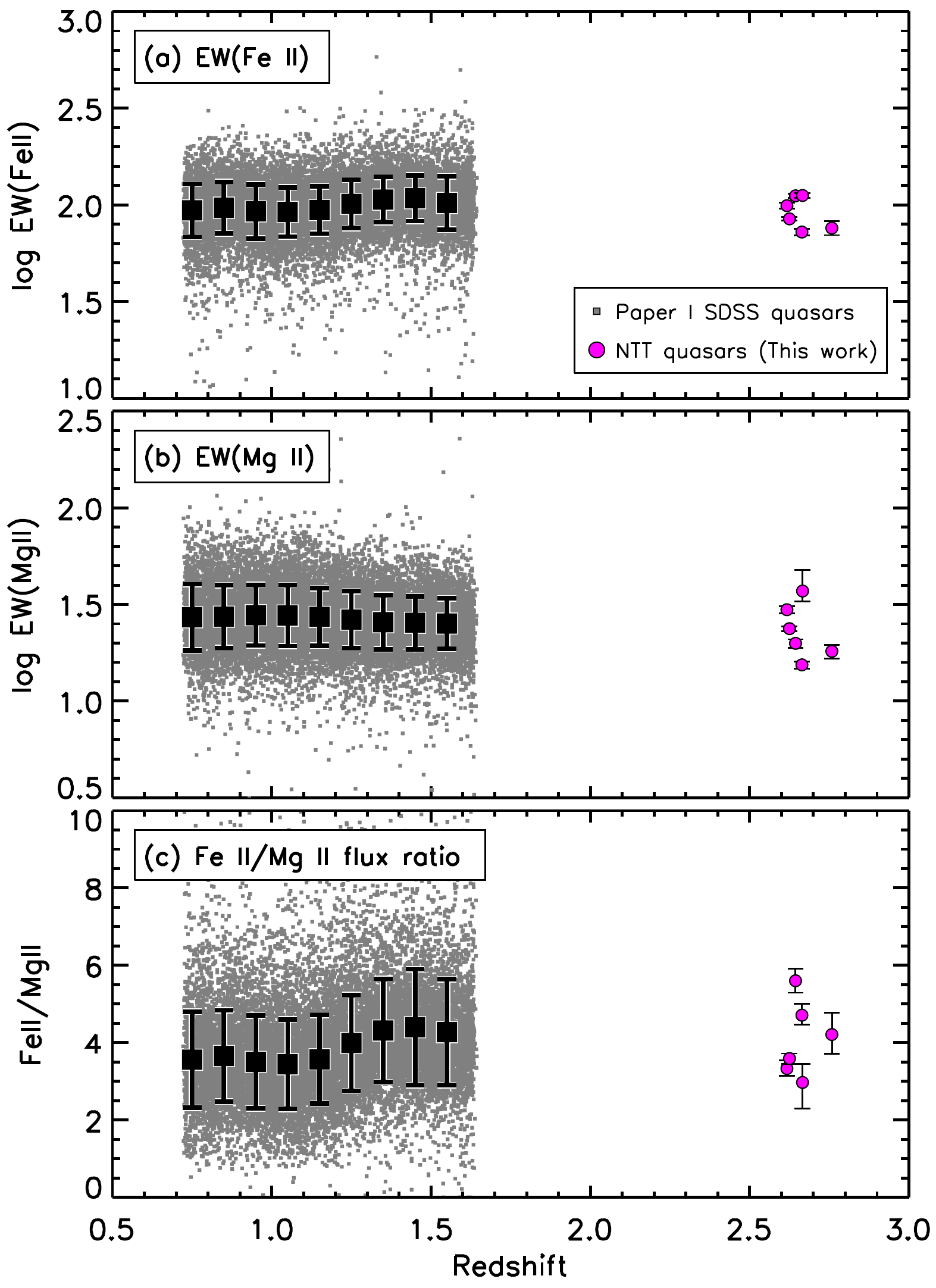}
 \caption{(a) EW(\ion{Fe}{2}), (b) EW(\ion{Mg}{2}), and (c)
 \ion{Fe}{2}/\ion{Mg}{2} flux ratio as a function of redshift for the
 six NTT quasars at $z \sim 2.7$ (magenta circle) and Paper~I SDSS
 quasars (gray dots).  For the latter data, the mean values and standard
 deviations (after 3$\sigma$ clipping) in each redshift bin ($\Delta z
 = 0.15$) are represented by filled boxes and error bars, respectively.
 }
 \label{fig:z_feiimgii}
\end{figure}

Figure \ref{fig:z_feiimgii} shows the EW(\ion{Fe}{2}), EW(\ion{Mg}{2}),
and \ion{Fe}{2}/\ion{Mg}{2} flux ratio\footnote{Here, the flux ratio
\ion{Fe}{2}/\ion{Mg}{2} equals the equivalent-width ratio
EW(\ion{Fe}{2})/EW(\ion{Mg}{2}), because we used the common continuum
flux at 3000~\AA\ for calculating EW(\ion{Fe}{2}) and EW(\ion{Mg}{2}) as
described in \S\ref{sec:analysis}.} as a function of redshift, along
with the Paper I SDSS quasars as reference.  Note that the EWs were
measured in their rest frames.  For all the three parameters, the
dynamic range of the measured values approximately match for the two
samples, despite the large difference in luminosity.  This suggests that
there is no evolution of the \ion{Fe}{2}/\ion{Mg}{2} flux ratio over a
long period of cosmic time, which is consistent with the results and
observations of previous studies (e.g.,
\citealt{2011ApJ...739...56D,2014ApJ...790..145D,2017ApJ...849...91M,2019ApJ...874...22S}).

\section{Discussion} \label{sec:discussion}

\subsection{Baldwin effect and metallicity} \label{sec:BEff}

\cite{1977ApJ...214..679B} found that the EW of \ion{C}{4} $\lambda1549$
emission line is anticorrelated with continuum luminosity in AGNs, which
is now called the BEff.  A similar anticorrelation was soon found for
\ion{Mg}{2} $\lambda2798$ (\citealt{1978Natur.273..431B}).  Currently,
this effect has been detected for almost all UV and optical emission
lines of AGNs, including \ion{Mg}{2} and \ion{Fe}{2} (e.g.,
\citealt{1995ApJS...99....1L,1999ASPC..162..235O,2002ApJ...581..912D,2007ASPC..373..355S}).
An intriguing aspect reported by \cite{2002ApJ...581..912D} is that the
slope, measured as $\mathrm{EW} \propto (\lambda L_\lambda)^\beta$,
becomes steeper as the ionization energy of the line increases.
Further, \cite{2008MNRAS.389.1703X} reported that the slope index
$\beta$ for \ion{C}{4} $\lambda1549$ seems to be independent of the
redshift up to $z \sim 5$.

While there is no doubt that the BEff exists from an observational
perspective, its physical origin remains an open question.  Several
scenarios have been proposed for this so far, such as (1) a
luminosity-dependent ionization parameter
(\citealt{1984ApJ...278..558M}), (2) an orientation-dependent
anisotropic continuum with a more isotropic line emission
(\citealt{1985MNRAS.216...63N}), and (3) a luminosity-dependent
continuum spectral energy distribution (SED;
\citealt{1998ApJ...507...24K,2002ApJ...581..912D}).

Given that the BEff has been detected even for hydrogen and helium lines
(e.g., \citealt{1995ApJS...99....1L,2002ApJ...581..912D}), there is some
evidence to conclude that its main cause is not metallicity.  This
argument may be supported by observations of metallicity-sensitive
nitrogen emission lines.  \cite{2008ApJ...679..962J} found that
$\sim$1.1\% of their quasar sample retrieved from SDSS DR5 have strong
\ion{N}{5} $\lambda1240$, \ion{N}{4}] $\lambda1486$, and \ion{N}{3}]
$\lambda1750$ emission lines.  These quasars are very likely to have a
large nitrogen abundance.  However, they reported that the difference in
luminosity between these quasars with strong nitrogen emission lines and
normal quasars is too small to explain the BEff.

On the other hand, \cite{1998ApJ...507...24K} argued metallicity as a
second driver of the BEff based on the fact that the BEff is not seen in
\ion{N}{5} $\lambda1240$.  Their photoionization calculation showed that
the EWs of various emission lines, including \ion{N}{5}, are weaker for
softer ionizing continua.  This result indicates that the luminosity
dependence of the SED, which becomes softer as the object becomes
brighter, is the primary driver of the BEff.  Then,
\cite{1998ApJ...507...24K} argued that if metallicity in addition to the
softness of the SED increases as the luminosity increases, the decrease
of EW(\ion{N}{5}) due to the SED can be compensated by the rapid
increase of \ion{N}{5} due to metallicity, which can reproduce the
observations.  \cite{2003ApJ...596...72W,2004ApJ...608..136W} showed
that the BEff can be reproduced by using the BH mass instead of
luminosity.  They also estimated the metallicity from line ratios
including \ion{N}{5} following the method of \cite{2002ApJ...564..592H}
and showed that there is a relationship between the BH mass and BLR
metallicity.  This may suggest that the BEff is related to the
mass-metallicity relationship found in galaxies.  However, it should be
noted that these results were obtained based on \ion{N}{5} and do not
necessarily agree with the results of other nitrogen emission lines.
\cite{2002ApJ...581..912D} detected the BEff for \ion{N}{3}]
$\lambda1750$ and \ion{N}{4}] $\lambda1486$, unlike \ion{N}{5}.
\cite{2003ApJ...596...72W} found no trend between the BH mass and
metallicity when \ion{N}{3}] instead of \ion{N}{5} was used as an
indicator of metallicity.

Given this indeterminacy of different results for different nitrogen
emission lines as well as the plausibility that the luminosity-dependent
SED is the primary driver, we consider the BEff to be independent of
chemical abundances in this study, and will correct measured EWs for the
BEff before abundance estimate.

\subsection{Baldwin effect for \ion{Mg}{2} and \ion{Fe}{2}}

\begin{figure}[t]
 \epsscale{1.1}
 \plotone{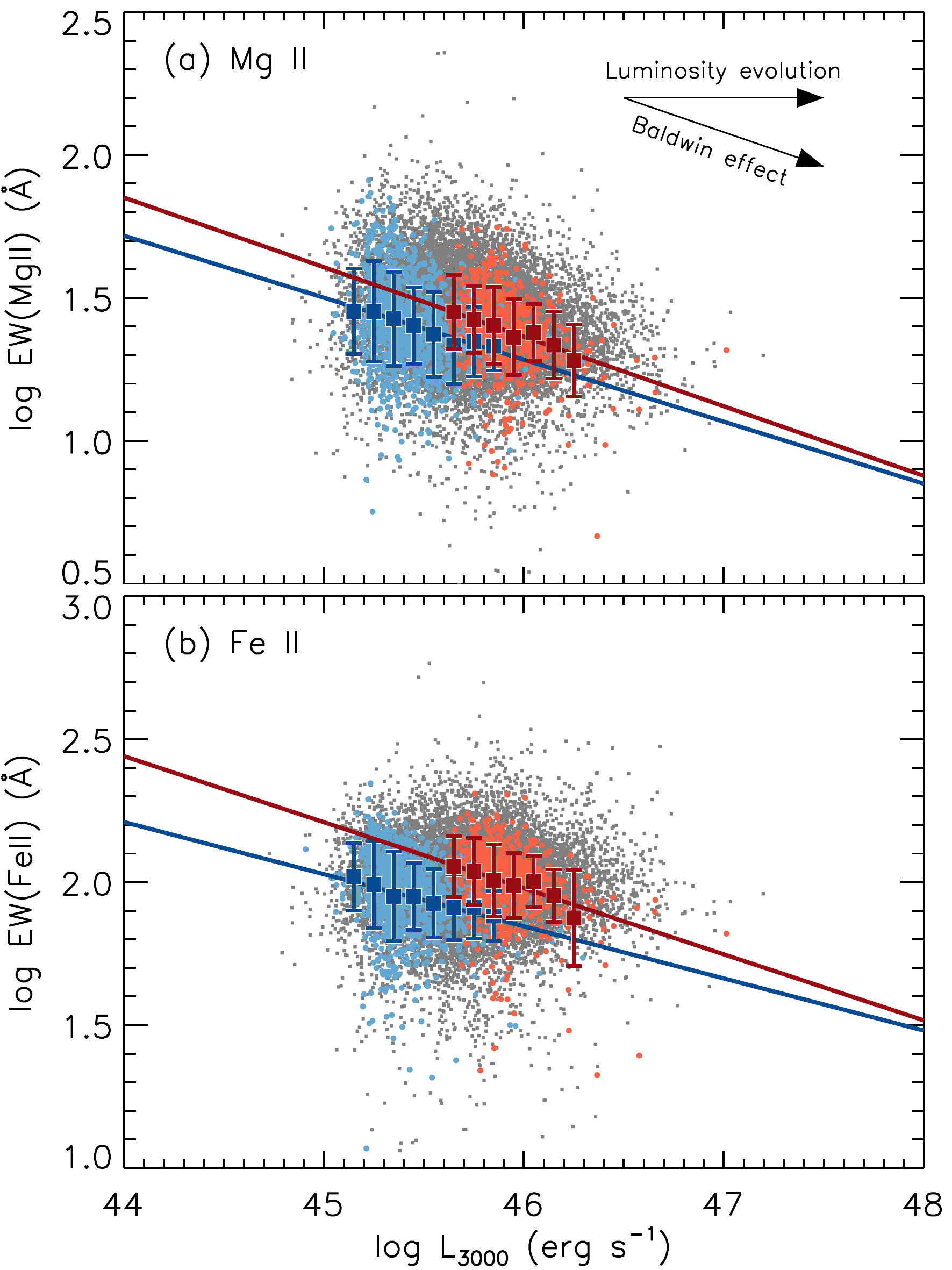}
 \caption{Distribution of Paper~I data on the EW--$\log L_{3000}$ plane
 for (a) \ion{Mg}{2} and (b) \ion{Fe}{2}.  Gray dots are Paper~I data
 with $\chi_\nu^2 < 1.2$ at $0.72 < z < 1.63$.  Redshift-binned
 subsamples ($0.8 < z < 0.9$ with blue circles; $1.4 < z < 1.5$ with red
 circles) are also shown.  The solid lines indicate regression lines of
 the two subsamples.}
 \label{fig:ew_lumi}
\end{figure}

\begin{deluxetable*}{lrrlrrlr}[t]
\tabletypesize{\footnotesize}
\tablecaption{Correlation between EW and $L_{3000}$ \label{tab:ew_lumi}}
\tablehead{
 \colhead{Redshift Range} & \multicolumn{3}{c}{EW(\ion{Mg}{2})} &
 \multicolumn{3}{c}{EW(\ion{Fe}{2})} & \colhead{$N_\mathrm{obj}$} \\
 \cmidrule(r){2-4} \cmidrule(r){5-7}
 & \colhead{Slope} & \colhead{$\rho$\tablenotemark{$\dagger$}} &
 \colhead{$p$\tablenotemark{$\ddagger$}} & \colhead{Slope} &
 \colhead{$\rho$\tablenotemark{$\dagger$}} & \colhead{$p$\tablenotemark{$\ddagger$}}
 }
 \startdata
 \multicolumn{8}{c}{Redshift-binned subsamples} \\
 \hline
 $0.7 < z < 0.8$   & $-0.17$ & $-0.17$ & $5\times10^{-7}$  & $-0.12$ & $-0.15$ & $1\times10^{-5}$  &  851 \\
 $0.8 < z < 0.9$   & $-0.22$ & $-0.25$ & $3\times10^{-17}$ & $-0.18$ & $-0.27$ & $5\times10^{-20}$ & 1089 \\
 $0.9 < z < 1.0$   & $-0.20$ & $-0.20$ & $2\times10^{-10}$ & $-0.22$ & $-0.29$ & $8\times10^{-20}$ &  971 \\
 $1.0 < z < 1.1$   & $-0.26$ & $-0.26$ & $7\times10^{-17}$ & $-0.19$ & $-0.27$ & $4\times10^{-18}$ & 1016 \\
 $1.1 < z < 1.2$   & $-0.27$ & $-0.28$ & $1\times10^{-16}$ & $-0.21$ & $-0.28$ & $3\times10^{-16}$ &  853 \\
 $1.2 < z < 1.3$   & $-0.29$ & $-0.32$ & $5\times10^{-19}$ & $-0.25$ & $-0.29$ & $3\times10^{-16}$ &  746 \\
 $1.3 < z < 1.4$   & $-0.21$ & $-0.23$ & $9\times10^{-10}$ & $-0.20$ & $-0.26$ & $9\times10^{-12}$ &  678 \\
 $1.4 < z < 1.5$   & $-0.24$ & $-0.28$ & $8\times10^{-13}$ & $-0.23$ & $-0.29$ & $5\times10^{-14}$ &  648 \\
 $1.5 < z < 1.6$   & $-0.28$ & $-0.31$ & $3\times10^{-16}$ & $-0.18$ & $-0.24$ & $2\times10^{-10}$ &  662 \\
 \hline
 \multicolumn{8}{c}{All sample} \\
 \hline
 $0.72 < z < 1.63$ & $-0.13$ & $-0.21$ & $<1\times10^{-30}$  & $-0.02$ & $-0.03$ & $0.02$ & 7710 \\
 \enddata

 \tablenotetext{\dagger}{Spearman's rank correlation coefficient.}
 \tablenotetext{\ddagger}{$p$-value for a null hypothesis that there is no correlation.}
\end{deluxetable*}

Here, we investigate the BEff for \ion{Mg}{2} and \ion{Fe}{2} emission
lines using the Paper~I SDSS quasars, which have a sufficient sample
size.  Figure \ref{fig:ew_lumi} shows the distribution of the Paper~I
SDSS quasars in the EW--$L_{3000}$ diagram for \ion{Mg}{2} and
\ion{Fe}{2}.  Because the Paper~I SDSS quasars are basically
flux-limited ($i=19.1$ for $z \lesssim 3$ quasars; see
\citealt{2010AJ....139.2360S}) and the limiting luminosity accordingly
depends on the redshift, we divided them into redshift-binned
subsamples.  As an example, two redshift-binned subsamples of $0.8 < z <
0.9$ and $1.4 < z < 1.5$ are plotted in different colors for clarity in
Figure \ref{fig:ew_lumi}.  Note that quasars at lower luminosity are
missing because of the flux limit of observation.  Table
\ref{tab:ew_lumi} summarizes the correlation analysis results.  Although
the luminosity range of each redshift-binned subsamples is narrow
($\sim$0.5~dex), anticorrelation was confirmed with the $p$-value $\le
10^{-5}$ for each subsample for both \ion{Mg}{2} and \ion{Fe}{2}.  The
slope index $\beta$ was in the range of $-0.17$ to $-0.29$ with $\langle
\beta \rangle = -0.24$ for \ion{Mg}{2}, while $-0.12$ to $-0.25$ with
$\langle \beta \rangle = -0.21$ for \ion{Fe}{2}.  A relatively large
scatter was observed for the values of $\beta$; however, no particular
redshift trend was observed, which is consistent with the result for
\ion{C}{4} reported by \cite{2008MNRAS.389.1703X}.

In Figure \ref{fig:ew_lumi}, a horizontal shift of the distribution can
be observed between the two redshift-binned subsamples.  As studies on
the luminosity function indicate, quasars show a drastic luminosity
evolution at $z \lesssim 2$ (e.g.,
\citealt{1988MNRAS.235..935B,2009MNRAS.399.1755C,2013ApJ...773...14R}),
which results in this horizontal shift.  When subsamples with the same
slope index $\beta$ are shifted horizontally, the slope of the whole
sample becomes shallow.  As a consequence, the BEff may be weakened or
hidden by the luminosity evolution of quasars when a whole sample with a
wide range of redshift is used to plot an EW--$L_{3000}$ diagram.  When
Paper~I SDSS quasars were used regardless of the redshift, we found
little or no correlation between the EW and the luminosity for
\ion{Fe}{2} (see Table \ref{tab:ew_lumi}).  As another example,
\cite{2008MNRAS.389.1703X} reported that the slope of the \ion{C}{4}
BEff is steeper for their redshift-binned subsample than for the whole
sample; this is naturally expected if quasars of whole samples covering
a wide range of redshift ($1.5 \lesssim z \lesssim 5.1$) underwent
luminosity evolution.  Thus, it is important to make subsamples in the
redshift to separate the luminosity evolution effect from the BEff.

\subsection{Relationship between the dependence on the Eddington ratio and the Baldwin effect} \label{sec:Edd_Beff}

As discussed in Paper~I, the EWs of \ion{Mg}{2} and \ion{Fe}{2} are
correlated with the Eddington ratio, which is a parameter that mainly
affects the observable properties of quasars
(\citealt{2002ApJ...565...78B}).  The following question now arises:
does the BEff possibly arise due to the dependence of the EW on the
Eddington ratio?  \cite{2009ApJ...703L...1D} investigated this point for
\ion{Mg}{2}.  They analyzed the spectroscopic data of Seyfert 1 galaxies
and quasars at $0.45 \le z \le 0.8$ retrieved from SDSS DR4, and they
found a strong negative correlation between EW(\ion{Mg}{2}) and
$L/L_\mathrm{Edd}$.  Furthermore, they found no correlation between
EW(\ion{Mg}{2}) and luminosity for their sample for the same
$L/L_\mathrm{Edd}$.  Therefore, they concluded that the BEff is purely
the secondary effect of the EW--$L/L_\mathrm{Edd}$ relationship.  Here,
we extend their study to include higher redshift data and investigate
both \ion{Mg}{2} and \ion{Fe}{2}.

We performed a two-dimensional least-squares fitting in the log scale,
where $L/L_\mathrm{Edd}$ and $L_{3000}$ were simultaneously taken as
independent variables, i.e.,
\begin{equation}
 \log \mathrm{EW} = \alpha \log \left( \frac{L/L_\mathrm{Edd}}{A}
		        \right) + \beta \log \left(
		        \frac{L_{3000}}{B} \right) +
		        \gamma, \label{eq:fmodel}
\end{equation}
where $\alpha$ and $\beta$ are the slope indices of the dependence on
the Eddington ratio and the BEff, respectively, and $A$ and $B$ are the
fiducial values of the independent variables.  Given that both the
dependence on the Eddington ratio and the BEff are independent to the
chemical composition of the gas, the constant $\gamma$ is expected to
strongly reflect the abundance.

In this study, we adopted the typical measured values for quasars as the
fiducial values $A$ and $B$, taking the following abundance estimate
into account.  As the dependence of the Eddington ratio on redshift is
only slight at $0.7 \lesssim z \lesssim 1.6$ (see
Figure~\ref{fig:luminosity_redshift}), we adopted a constant value as
the fiducial value of the Eddington ratio:
\begin{equation}
 A = 10^{-0.55}, \label{eq:edd_fid}
\end{equation}
which was retrieved from the median value of the whole Paper~I SDSS
quasars, as was done in Paper~I.

To determine the fiducial value of luminosity $B$, we referred to
studies on the quasar luminosity function.  \cite{2013ApJ...773...14R}
measured the luminosity functions of quasars using data from the
SDSS-III Baryon Oscillation Spectroscopic Survey; they found that the
measured luminosity functions can be fitted with a pure luminosity
evolution model at $0.3 < z < 2.2$, whereas a luminosity evolution
density evolution model is needed at $2.2 < z < 3.5$.  Following their
parameterization, the characteristic (or break) absolute magnitude is
written as
\begin{equation}
 M_i^*(z) = \begin{cases}
	 M_i^*(0) - 2.5 (k_1 z - k_2 z^2) & (0.3 < z < 2.2) \\
	 M_i^*(2.2) + c_2 (z - 2.2) & (2.2 < z < 3.5),
	\end{cases}
\end{equation}
where $M_i^*(z)$ is the absolute magnitude at which the faint-end and
bright-end slopes of a luminosity function at redshift $z$ overlap,
$k_1=1.241$, $k_2=-0.249$, and $c_2=-0.875$.  This equation can be
written equivalently in terms of luminosity as
\begin{equation}
 \log L^*(z) = \begin{cases}
	    \log L^*(0) + k_1 z - k_2 z^2 & (0.3 < z < 2.2) \\
	    \log L^*(2.2) - (c_2 / 2.5) (z - 2.2) & (2.2 < z < 3.5). \label{eq:lumi_evo}
          \end{cases}
\end{equation}
Assuming that typical monochromatic luminosity $\langle L_{3000}
\rangle$ also changes with redshift as in Equation (\ref{eq:lumi_evo}),
we determine the zero-point by minimizing $\chi_\nu^2$ for the Paper~I
SDSS quasars and derive $\langle L_{3000} \rangle(z=0)=10^{44.63}$
erg~s$^{-1}$.  The typical luminosity $\langle L_{3000} \rangle(z)$ thus
determined is plotted as a function of redshift in Figure
\ref{fig:luminosity_redshift}; it is evident that $\langle L_{3000}
\rangle(z)$ agrees with the distribution of Paper~I data very well and
is suitable for the fiducial luminosity.  Therefore, we adopt $\langle
L_{3000} \rangle(z)$ as the fiducial value $B$ in
Equation~(\ref{eq:fmodel}), i.e.,
\begin{equation}
 B = \begin{cases}
      10^{44.63} + 1.241 z + 0.249 z^2 & (0.3 < z < 2.2) \\
      10^{46.15} + 0.350 (z - 2.2) & (2.2 < z < 3.5), \label{eq:lumi_fid}
     \end{cases}
\end{equation}
in the unit of erg~s$^{-1}$.

\begin{figure*}[t]
 \epsscale{1.1}
 \plotone{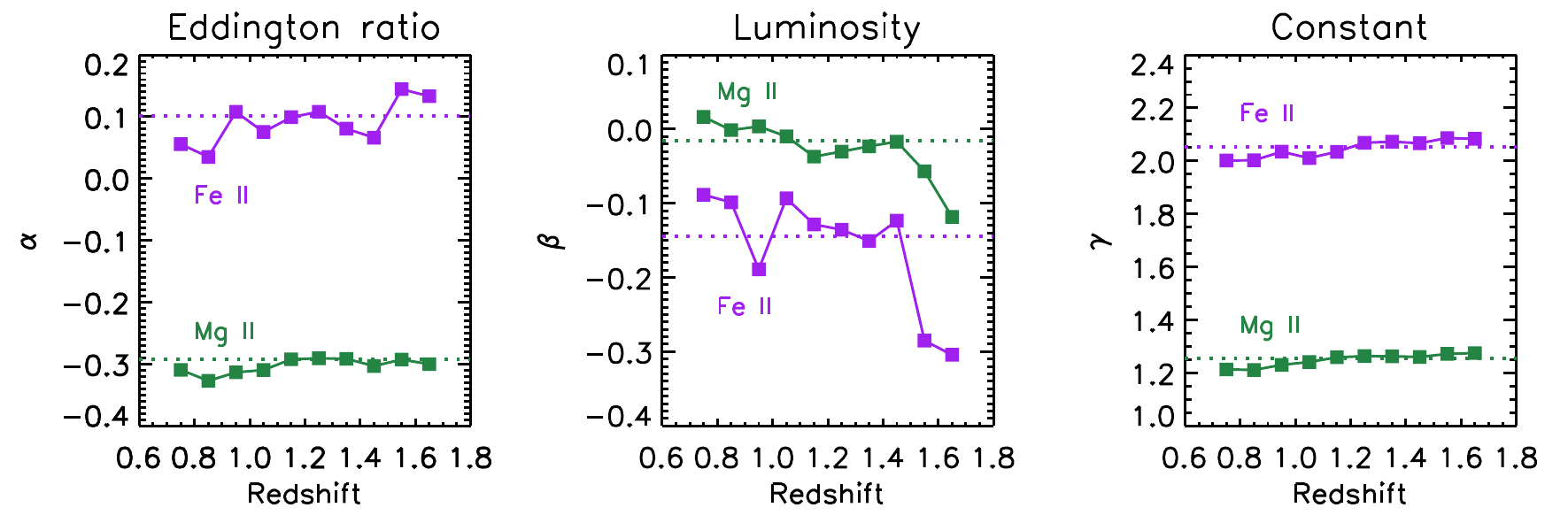}
 \caption{Parameters retrieved from the simultaneous fitting of EW with
 Equation~(\ref{eq:fmodel}) as a function of redshift for \ion{Mg}{2}
 (green) and \ion{Fe}{2} (purple).  Dashed lines represent the results
 of the fitting using the whole sample; $(\alpha, \beta, \gamma)=(-0.29,
 -0.02, +1.25)$ for \ion{Mg}{2}, and $(+0.10, -0.15, +2.05)$ for
 \ion{Fe}{2}.}
 \label{fig:planefit_lumievo}
\end{figure*}

By using Equation~(\ref{eq:fmodel}) with the above $A$ and $B$, we
performed a simultaneous fitting of EWs; the results are shown in
Figure~\ref{fig:planefit_lumievo}.  Notable characteristics of the
fitting are summarized as follows:

\begin{enumerate}
 \item $\alpha$ and $\beta$ clearly show different patterns between
       \ion{Mg}{2} and \ion{Fe}{2}, even though they have similar
       ionization potential.
 \item The slope index of the Eddington ratio $\alpha$ shows little or
       no dependence on redshift.  The values are almost consistent with
       the results of Paper~I for both \ion{Mg}{2} and \ion{Fe}{2}.
 \item The slope index of luminosity $\beta$ may depend on redshift.
       Generally, the higher the redshift, the steeper the slope for
       both \ion{Mg}{2} and \ion{Fe}{2}.
 \item $\gamma$ is approximately constant for redshifts for both
       \ion{Mg}{2} and \ion{Fe}{2}.  This is consistent with the
       theoretical expectations of the chemical evolution models that
       the chemical abundance of the gas was nearly constant at $0.7
       \lesssim z \lesssim 1.6$.
\end{enumerate}

Our result shows that $\beta$(\ion{Mg}{2}) is almost zero at $z \lesssim
1.5$, implying no correlation between EW(\ion{Mg}{2}) and luminosity,
which is consistent with the argument proposed by
\cite{2009ApJ...703L...1D}.  However, at $z \gtrsim 1.5$, which was not
investigated by \cite{2009ApJ...703L...1D}, $\beta$(\ion{Mg}{2}) becomes
nonzero.  Furthermore, $\beta$(\ion{Fe}{2}) is nonzero regardless of
redshift.  These results indicate that the BEff is more than a secondary
effect in the EW(\ion{Mg}{2})--$L/L_\mathrm{Edd}$ relationship.  The
different redshift dependences of $\alpha$ and $\beta$ shown in
Figure~\ref{fig:planefit_lumievo} also suggest that the two effects are
independent to some extent.  This idea may be supported by the principal
component analysis of the quasar spectra, which was performed by
\cite{2003ApJ...586...52S}; they found that the first principal
component represents the BEff, while the third principal component is
directly related to ``Eigenvector 1''
(\citealt{1992ApJS...80..109B,2000ApJ...536L...5S}), which is thought to
be mainly driven by the Eddington ratio
(\citealt{2002ApJ...565...78B})\footnote{For reference, the second
principal component derived by \cite{2003ApJ...586...52S} represents the
UV continuum slope variation in quasar spectra.}.

Further investigation of the difference between the BEff and the
dependence on the Eddington ratio may be helpful for solving
long-standing issues regarding the BEff and for understanding the
physical mechanisms of quasars.  These topics are beyond the scope of
this paper.

\subsection{Abundance estimate} \label{sec:abundance}

Given that there are observational results supporting that the BEff is
not related to metallicity (see \S\ref{sec:BEff}, e.g.,
\citealt{1995ApJS...99....1L,2002ApJ...581..912D}), in addition to
correcting for their dependence on the Eddington ratio, it is necessary
to correct the measured EWs for the BEff to estimate the chemical
abundance of a BLR cloud.  This is especially important for most
high-redshift quasars found thus far, including the NTT quasars analyzed
in this study, because they are extraordinarily luminous compared with
typical quasars at similar redshift.  Thus, their EWs could be smaller
than those of typical quasars due to the BEff.  Here, we attempted to
correct the measured EWs for the BEff in addition to the dependence on
the Eddington ratio and estimated their chemical abundances by using the
method described in Paper~I.

Following the discussion in the previous section, we can derive an EW
correction formula for the dependence on nonabundance factors: 
\begin{equation}
 \mathrm{EW}^\prime = \mathrm{EW} \left( \frac{L/L_\mathrm{Edd}}{A}
		         \right)^{-\alpha} \left(
		         \frac{L_{3000}}{B} \right)^{-\beta}, \label{eq:ew_cor}
\end{equation}
where EW is the measured EW, and $A$ and $B$ are given in
Equations~(\ref{eq:edd_fid}) and (\ref{eq:lumi_fid}), respectively.
Figure~\ref{fig:planefit_lumievo} implies that $\alpha$ does not depend
on redshift; therefore, we can safely adopt the value retrieved from the
fitting result using the whole sample in the previous section (see
Figure~\ref{fig:planefit_lumievo}), i.e., $\alpha=-0.29$ (\ion{Mg}{2})
and $+0.10$ (\ion{Fe}{2}).  In contrast, $\beta$ may depend on redshift;
however, sufficient data is not available for $z \gtrsim 1.6$.  In this
study, we considered the following two working hypotheses to derive the
chemical abundances of the NTT quasars:
\begin{enumerate}
 \item WH1: $\beta$ does not depend on redshift.  We adopt $\beta=-0.02$
       for \ion{Mg}{2} and $-0.15$ for \ion{Fe}{2} (see the caption of
       Figure~\ref{fig:planefit_lumievo}), which are retrieved from the
       fitting result using the whole sample in the same manner as
       $\alpha$.
 \item WH2: $\beta$ depends on redshift.  We adopt the same $\beta$ as
       WH1 at $z < 1.5$, but adopt $\beta=-0.10$ for \ion{Mg}{2} and
       $-0.30$ for \ion{Fe}{2} at $z \ge 1.5$, which are inferred from
       Figure~\ref{fig:planefit_lumievo}.
\end{enumerate}

\begin{figure}[t]
 \epsscale{1.1}
 \plotone{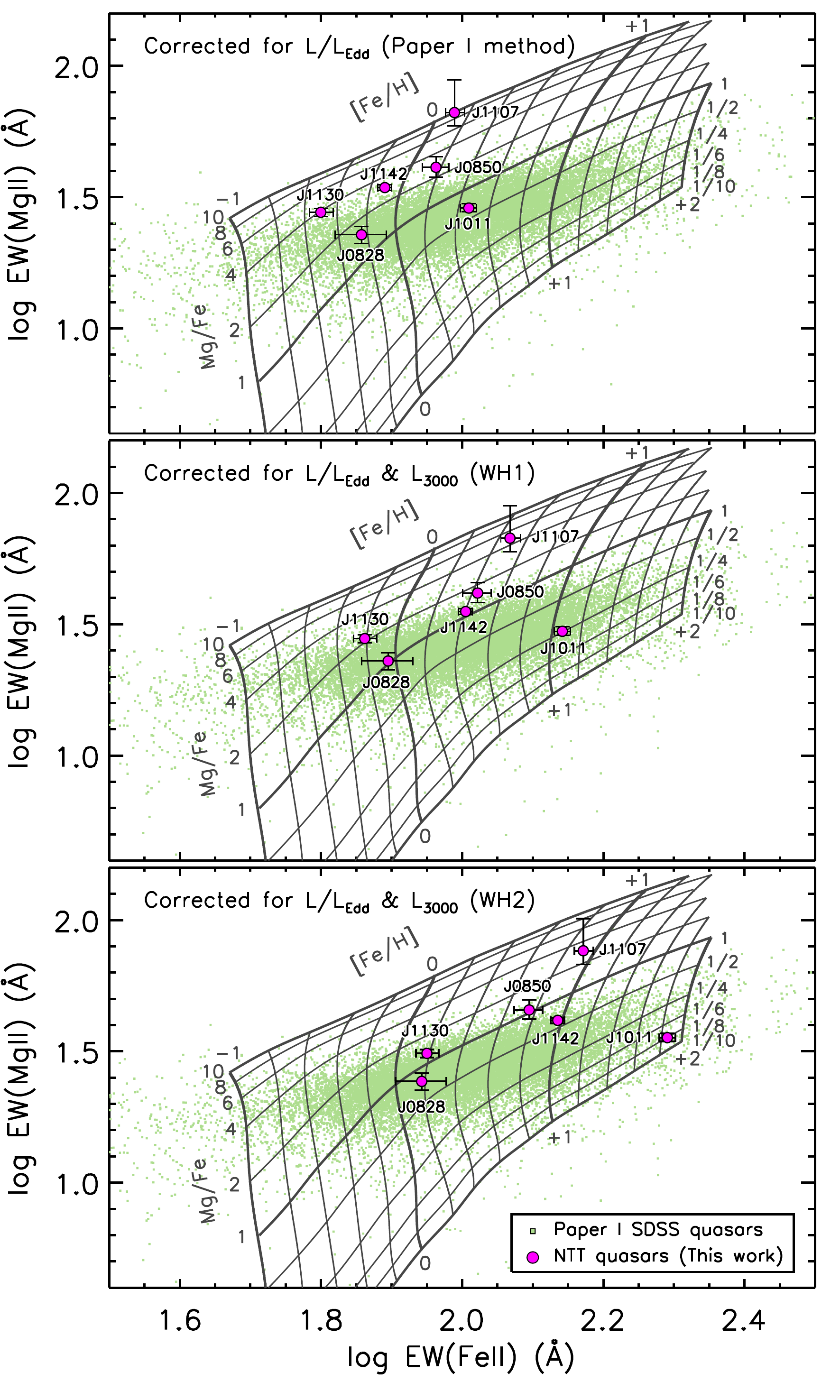}
 \caption{Calculated variation of EW(\ion{Mg}{2}) and EW(\ion{Fe}{2}) as
 functions of Mg/Fe ($= 10^\mathrm{[Mg/Fe]}$) and [Fe/H] (grid lines).
 The measured EWs corrected for only $L/L_\mathrm{Edd}$ (top panel), for
 both $L/L_\mathrm{Edd}$ and $L_{3000}$ with WH1 (middle panel), and
 that with WH2 (bottom panel) are plotted for NTT quasars (magenta
 circle) and Paper~I SDSS quasars (gray dot).}
 \label{fig:ewew_contour_cor}
\end{figure}

\begin{figure*}[t]
 \epsscale{1.1}
 \plotone{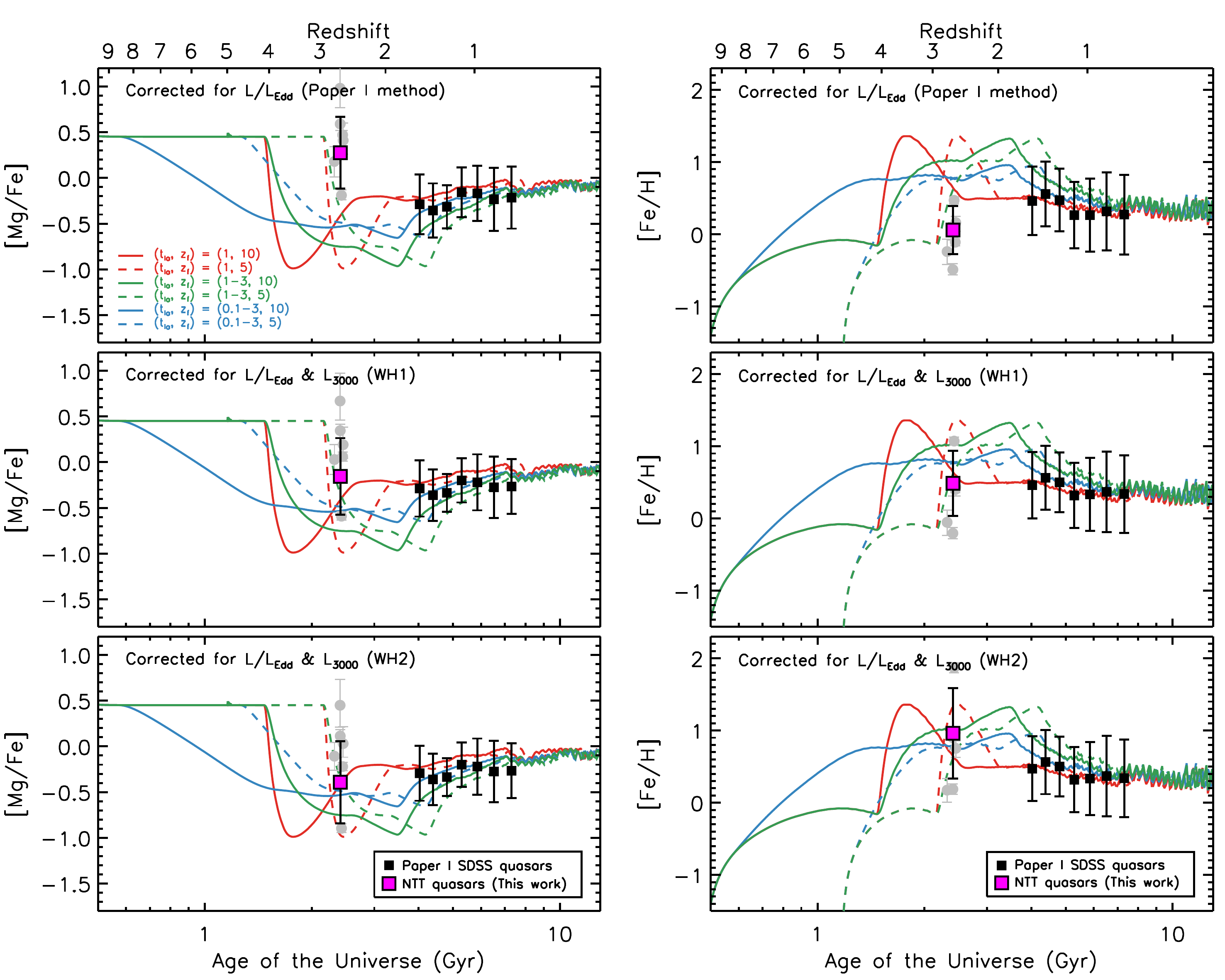}
 \caption{[Mg/Fe] (left panels) and [Fe/H] (right panels) as a function
 of the age of the universe.  The abundances are derived from the
 measured EWs corrected for only $L/L_\mathrm{Edd}$ (top panels) and for
 both $L/L_\mathrm{Edd}$ and $L$ with WH1 (middle panels) and WH2
 (bottom panels).  The overplotted curves are the chemical evolution
 models of quasar host galaxies for several prescriptions of the SNe Ia
 lifetime $t_\mathrm{Ia}$ in Gyr and the initial star formation redshift
 $z_f$ (see Paper~I).}
 \label{fig:mgfe_feh_age}
\end{figure*}

Figure~\ref{fig:ewew_contour_cor} shows the abundance diagnostic diagram
introduced in Paper~I, where the calculated variation in EW(\ion{Mg}{2})
and EW(\ion{Fe}{2}) as functions of [Mg/Fe] and [Fe/H] is compared with
the measured EWs of the NTT quasars and Paper~I SDSS quasars after
correction for the nonabundance factors.  To check the effect of
additional correction for the BEff, we have plotted three patterns:
measured EWs corrected for only $L/L_\mathrm{Edd}$ with the Paper~I
method (top panel), and those corrected for both $L/L_\mathrm{Edd}$ and
$L_{3000}$ using Equation~(\ref{eq:ew_cor}) with WH1 (middle panel) and
WH2 (bottom panel).  The derived [Mg/Fe] and [Fe/H] for these three
cases are shown as a function of the age of the universe in
Figure~\ref{fig:mgfe_feh_age}.  The numerical values for these cases are
listed in Table~\ref{tab:abundance}.  Because the sample size of the
Paper~I SDSS quasars is large, the mean and standard deviation are
calculated in each redshift bin ($\Delta z=0.15$) and plotted in this
figure.  Similarly, the six NTT quasars are treated as one group and
their weighted mean and standard deviation are plotted for proper
comparison.

\begin{deluxetable*}{lrrrrrr}[t]
\tablecaption{Derived abundances \label{tab:abundance}}
\tablehead{
 \colhead{Redshift Range} & \multicolumn{2}{c}{Corrected for
 $L/L_\mathrm{Edd}$} & \multicolumn{2}{c}{Corrected for
 $L/L_\mathrm{Edd}$ and $L_{3000}$} & \multicolumn{2}{c}{Corrected for
 $L/L_\mathrm{Edd}$ and $L_{3000}$} \\
  & \multicolumn{2}{c}{(Paper~I method)} & \multicolumn{2}{c}{(WH1)} & \multicolumn{2}{c}{(WH2)} \\
 \cmidrule(r){2-3} \cmidrule(r){4-5} \cmidrule(r){6-7}
 & \colhead{[Mg/Fe]} & \colhead{[Fe/H]} & \colhead{[Mg/Fe]} & \colhead{[Fe/H]} & \colhead{[Mg/Fe]} & \colhead{[Fe/H]}
 }
 \startdata
 \multicolumn{7}{c}{Paper~I SDSS quasars} \\
 \hline
 $0.60 < z < 0.75$ & $-0.22\pm0.34$ & $+0.27\pm0.55$ & $-0.27\pm0.30$ & $+0.34\pm0.54$ & $-0.27\pm0.30$ & $+0.34\pm0.54$ \\
 $0.75 < z < 0.90$ & $-0.23\pm0.34$ & $+0.32\pm0.54$ & $-0.27\pm0.34$ & $+0.37\pm0.56$ & $-0.27\pm0.34$ & $+0.37\pm0.56$ \\
 $0.90 < z < 1.05$ & $-0.17\pm0.30$ & $+0.26\pm0.51$ & $-0.22\pm0.31$ & $+0.33\pm0.51$ & $-0.22\pm0.31$ & $+0.33\pm0.51$ \\
 $1.05 < z < 1.20$ & $-0.16\pm0.27$ & $+0.26\pm0.46$ & $-0.20\pm0.24$ & $+0.32\pm0.45$ & $-0.20\pm0.24$ & $+0.32\pm0.45$ \\
 $1.20 < z < 1.35$ & $-0.32\pm0.24$ & $+0.47\pm0.43$ & $-0.34\pm0.21$ & $+0.50\pm0.41$ & $-0.34\pm0.21$ & $+0.50\pm0.41$ \\
 $1.35 < z < 1.50$ & $-0.36\pm0.30$ & $+0.56\pm0.45$ & $-0.36\pm0.28$ & $+0.56\pm0.44$ & $-0.36\pm0.28$ & $+0.56\pm0.44$ \\
 $1.50 < z < 1.65$ & $-0.29\pm0.33$ & $+0.46\pm0.47$ & $-0.29\pm0.30$ & $+0.46\pm0.46$ & $-0.29\pm0.30$ & $+0.47\pm0.45$ \\
 \hline
 \multicolumn{7}{c}{NTT quasars} \\
 \hline
 $2.61 < z < 2.76$ & $+0.28\pm0.39$ & $+0.06\pm0.33$ & $-0.16\pm0.42$ & $+0.49\pm0.45$ & $-0.39\pm0.45$ & $+0.96\pm0.62$ \\
 \enddata
\end{deluxetable*}

In Figure~\ref{fig:mgfe_feh_age}, the Paper~I SDSS quasars at $z < 2$
are almost unaffected by the additional correction for the BEff.  This
is because the luminosity difference from the fiducial value, written as
$\Delta \log L \equiv \log L_{3000} - \log \langle L_{3000} \rangle(z)$,
is less than $\pm0.25$ dex for 68\% of the sample; this changes the
EW(\ion{Fe}{2}) by less than $\pm10$\% with the assumption
$\beta=-0.30$.  In contrast, all the NTT quasars are more luminous than
$\langle L_{3000} \rangle(z)$ with the mean difference $\langle \Delta
\log L \rangle = +0.6$ dex; this changes the EW(\ion{Fe}{2}) by
approximately $-50$\% with the assumption $\beta = -0.30$.  Therefore,
the derived abundances are almost unaffected for the Paper~I SDSS
quasars.  This implies that the conclusions of Paper~I are unchanged.

Further, the result for the NTT quasars at $z \sim 2.7$ depends on
whether the measured EWs have been corrected properly for the
nonabundance parameters, especially at $z \gtrsim 1.5$ where the slope
change of the BEff is shown in the middle panel of
Figure~\ref{fig:planefit_lumievo}.  When only the dependence on the
Eddington ratio is corrected using the Paper~I method (top panels of
Figure~\ref{fig:mgfe_feh_age}), the derived [Mg/Fe] at $z \sim 2.7$ is
greater than that obtained in Paper~I, while the derived [Fe/H] at $z
\sim 2.7$ is smaller than that in Paper~I.  The comparison of these
results with those of the chemical evolution models
(\citealt{1996ApJ...462..266Y,1998ApJ...507L.113Y}) shows that the
initial star formation redshift of $z_f=5$ is required to reproduce the
observation, which is clearly inconsistent with the existence of quasars
at $z > 5$.  In contrast, when additional correction is applied for the
BEff, the derived [Mg/Fe] and [Fe/H] become more consistent with the
models for $z_f = 10$ (middle panels of Figure~\ref{fig:mgfe_feh_age}),
which are considered more appropriate according to our current knowledge
about most distant objects in the universe.  In particular, if we adopt
WH2 (bottom panels of Figure~\ref{fig:mgfe_feh_age}), the derived
[Mg/Fe] and [Fe/H] agree well with the models for $z_f =10$.  Although
it is too early to put constraints on $z_f$ given the limited number of
the NTT quasars, the above results imply the following: (1) correction
for the BEff in addition to the dependence on the Eddington ratio is
needed to derive Mg and Fe abundances and their ratio from the measured
EWs of \ion{Mg}{2} and \ion{Fe}{2} emission lines, and (2) the slope
index of the BEff after separating the correction for the Eddington
ratio likely evolves with redshift.

Our findings will be important for extending the chemical evolution
studies using EWs of \ion{Mg}{2} and \ion{Fe}{2} emission lines of most
distance quasars.  However, further verification of our working
hypotheses, i.e., whether the correlation slope of the BEff for
\ion{Mg}{2} and \ion{Fe}{2} emission lines evolves even at $z \gtrsim 2$
is essential.  This investigation will require a large sample of
high-redshift quasars with sufficient dynamic range of luminosity.  This
will be implemented by using existing large telescopes (Subaru, VLT,
Keck, etc.) as well as an upcoming 6.5~m infrared-optimized telescope in
Chile constructed by The University of Tokyo Atacama Observatory (TAO)
project
(\citealt{2002aprm.conf...35Y,2014SPIE.9145E..07Y,2016SPIE.9906E..0RY,2018SPIE10700E..0WD}).


\section{Summary} \label{sec:summary}

We performed near-infrared spectroscopy for luminous six quasars at $z
\sim 2.7$ with the WINERED spectrograph mounted on the NTT at the La
Silla Observatory, Chile.  By combining these spectra with the optical
spectra obtained by SDSS, we measured the EWs of \ion{Mg}{2} and
\ion{Fe}{2} emission lines using the same method as in Paper~I
(\citealt{2017ApJ...834..203S}).  The measured \ion{Fe}{2}/\ion{Mg}{2}
flux ratios showed no significant change compared with those at $0.7
\lesssim z \lesssim 1.6$ (Figure~\ref{fig:z_feiimgii}), which is
consistent with the results of previous studies, which argued that there
was no redshift evolution of the \ion{Fe}{2}/\ion{Mg}{2} flux ratio.

From the analysis of Paper~I SDSS quasars, we found the BEff, i.e., EW
and luminosity are anticorrelated for both \ion{Mg}{2} and \ion{Fe}{2}
emission lines (Figure~\ref{fig:ew_lumi}).  The two-dimensional
least-squares fitting of measured EWs, which incorporates the effect of
quasar evolution inferred from a recent study on quasar luminosity
function, indicated that the BEff and the dependence on the Eddington
ratio are independent to some extent.  The correlation slope index
$\beta$ of the BEff showed the feature of redshift evolution
(Figure~\ref{fig:planefit_lumievo}).

Based on previous studies supporting that the BEff does not represent
metallicity, we corrected the measured EWs for the BEff in addition to
the dependence on the Eddington ratio, which was done in Paper~I.  For
the correction of EWs for the BEff, we assumed two working hypotheses
that the slope index $\beta$ does not evolve or evolves with redshift.
By comparing these corrected EWs with those calculated through
photoionization simulations in which the chemical abundance pattern of a
BLR cloud was varied, we derived [Mg/Fe] and [Fe/H] for the NTT quasars
and Paper~I SDSS quasars (Table~\ref{tab:abundance}).  We found that
both the derived [Mg/Fe] and [Fe/H] were consistent with the prediction
of chemical evolution models at $0.7 \lesssim z \lesssim 2.7$ when the
additional correction for the BEff with redshift-dependent $\beta$ is
applied (Figure~\ref{fig:mgfe_feh_age}).  This validates our method of
flux-to-abundance conversion.

In the future, increasing the sample size of high-redshift quasars with
sufficient dynamic range of luminosity is important for investigating
the BEff at high redshift.  This will be implemented by using large
telescopes including an upcoming 6.5 m infrared-optimized telescope
developed by the TAO project.


\acknowledgments
\vfill

We acknowledge the anonymous referee for constructive comments that
helped improve the quality of our manuscript.  We are grateful to the
staff of the Koyama Astronomical Observatory for their support during our
observation.  We also thank Takuji Tsujimoto and Masao Mori for useful
discussion.  H.S. is supported by the Japan Society for the Promotion of
Science (JSPS) KAKENHI grant No. 19K03917.  The travel expenses to
perform observation at NTT were supported by the Hayakawa Satio Fund in
the Astronomical Society of Japan.  WINERED was developed by the
University of Tokyo and the Laboratory of Infrared High-resolution
spectroscopy (LiH), Kyoto Sangyo University under the financial supports
of Grants-in-Aid, KAKENHI, from JSPS (Nos. 16684001, 20340042, and
21840052) and the MEXT Supported Program for the Strategic Research
Foundation at Private Universities (Nos. S081061 and S1411028).  N.K. is
supported by JSPS-DST under the Japan-India Science Cooperative Programs
during 2013--2015 and 2016--2018.  S.H. is supported by Grant-in-Aid for
JSPS Fellows grant No. 13J10504.  M.M. is supported by the Hakubi
project at Kyoto University.  K.F. is supported by KAKENHI (16H07323)
Grant-in-Aid for Research Activity start-up.

Funding for SDSS-III has been provided by the Alfred P. Sloan
Foundation, the Participating Institutions, the National Science
Foundation, and the U.S. Department of Energy Office of Science. The
SDSS-III website is http://www.sdss3.org/.

SDSS-III is managed by the Astrophysical Research Consortium for the
Participating Institutions of the SDSS-III Collaboration including the
University of Arizona, the Brazilian Participation Group, Brookhaven
National Laboratory, Carnegie Mellon University, University of Florida,
the French Participation Group, the German Participation Group, Harvard
University, the Instituto de Astrofisica de Canarias, the Michigan
State/Notre Dame/JINA Participation Group, Johns Hopkins University,
Lawrence Berkeley National Laboratory, Max Planck Institute for
Astrophysics, Max Planck Institute for Extraterrestrial Physics, New
Mexico State University, New York University, Ohio State University,
Pennsylvania State University, University of Portsmouth, Princeton
University, the Spanish Participation Group, University of Tokyo,
University of Utah, Vanderbilt University, University of Virginia,
University of Washington, and Yale University.




\begin{thebibliography}{}
\expandafter\ifx\csname natexlab\endcsname\relax\def\natexlab#1{#1}\fi

\bibitem[{{Baldwin}(1977)}]{1977ApJ...214..679B}
{Baldwin}, J.~A. 1977, \apj, 214, 679

\bibitem[{{Baldwin} {et~al.}(1978){Baldwin}, {Burke}, {Gaskell}, \&
  {Wampler}}]{1978Natur.273..431B}
{Baldwin}, J.~A., {Burke}, W.~L., {Gaskell}, C.~M., \& {Wampler}, E.~J. 1978,
  \nat, 273, 431

\bibitem[{{Barth} {et~al.}(2003){Barth}, {Martini}, {Nelson}, \&
  {Ho}}]{2003ApJ...594L..95B}
{Barth}, A.~J., {Martini}, P., {Nelson}, C.~H., \& {Ho}, L.~C. 2003, \apjl,
  594, L95

\bibitem[{{Boroson}(2002)}]{2002ApJ...565...78B}
{Boroson}, T.~A. 2002, \apj, 565, 78

\bibitem[{{Boroson} \& {Green}(1992)}]{1992ApJS...80..109B}
{Boroson}, T.~A., \& {Green}, R.~F. 1992, \apjs, 80, 109

\bibitem[{{Boyle} {et~al.}(1988){Boyle}, {Shanks}, \&
  {Peterson}}]{1988MNRAS.235..935B}
{Boyle}, B.~J., {Shanks}, T., \& {Peterson}, B.~A. 1988, \mnras, 235, 935

\bibitem[{{Croom} {et~al.}(2009){Croom}, {Richards}, {Shanks}, {Boyle},
  {Strauss}, {Myers}, {Nichol}, {Pimbblet}, {Ross}, {Schneider}, {Sharp}, \&
  {Wake}}]{2009MNRAS.399.1755C}
{Croom}, S.~M., {Richards}, G.~T., {Shanks}, T., {et~al.} 2009, \mnras, 399,
  1755

\bibitem[{{Cutri} {et~al.}(2003){Cutri}, {Skrutskie}, {van Dyk}, {Beichman},
  {Carpenter}, {Chester}, {Cambresy}, {Evans}, {Fowler}, {Gizis}, {Howard},
  {Huchra}, {Jarrett}, {Kopan}, {Kirkpatrick}, {Light}, {Marsh}, {McCallon},
  {Schneider}, {Stiening}, {Sykes}, {Weinberg}, {Wheaton}, {Wheelock}, \&
  {Zacarias}}]{2003tmc..book.....C}
{Cutri}, R.~M., {Skrutskie}, M.~F., {van Dyk}, S., {et~al.} 2003, {2MASS All
  Sky Catalog of point sources.}

\bibitem[{{De Rosa} {et~al.}(2011){De Rosa}, {Decarli}, {Walter}, {Fan},
  {Jiang}, {Kurk}, {Pasquali}, \& {Rix}}]{2011ApJ...739...56D}
{De Rosa}, G., {Decarli}, R., {Walter}, F., {et~al.} 2011, \apj, 739, 56

\bibitem[{{De Rosa} {et~al.}(2014){De Rosa}, {Venemans}, {Decarli}, {Gennaro},
  {Simcoe}, {Dietrich}, {Peterson}, {Walter}, {Frank}, {McMahon}, {Hewett},
  {Mortlock}, \& {Simpson}}]{2014ApJ...790..145D}
{De Rosa}, G., {Venemans}, B.~P., {Decarli}, R., {et~al.} 2014, \apj, 790, 145

\bibitem[{{Dietrich} {et~al.}(2002{\natexlab{a}}){Dietrich}, {Appenzeller},
  {Vestergaard}, \& {Wagner}}]{2002ApJ...564..581D}
{Dietrich}, M., {Appenzeller}, I., {Vestergaard}, M., \& {Wagner}, S.~J.
  2002{\natexlab{a}}, \apj, 564, 581

\bibitem[{{Dietrich} {et~al.}(2003){Dietrich}, {Hamann}, {Appenzeller}, \&
  {Vestergaard}}]{2003ApJ...596..817D}
{Dietrich}, M., {Hamann}, F., {Appenzeller}, I., \& {Vestergaard}, M. 2003,
  \apj, 596, 817

\bibitem[{{Dietrich} {et~al.}(2002{\natexlab{b}}){Dietrich}, {Hamann},
  {Shields}, {Constantin}, {Vestergaard}, {Chaffee}, {Foltz}, \&
  {Junkkarinen}}]{2002ApJ...581..912D}
{Dietrich}, M., {Hamann}, F., {Shields}, J.~C., {et~al.} 2002{\natexlab{b}},
  \apj, 581, 912

\bibitem[{{Doi} {et~al.}(2018){Doi}, {Miyata}, {Yoshii}, {Kohno}, {Tanaka},
  {Motohara}, {Minezaki}, {Kawara}, {Sako}, {Morokuma}, {Tamura}, {Tanabe},
  {Hatsukade}, {Takahashi}, {Konishi}, {Kamizuka}, {Kato}, {Aoki}, {Soyano},
  {Tarusawa}, {Handa}, {Koshida}, {Bronfman}, {Ruiz}, {Hamuy}, {Mendez},
  {Garay}, \& {Escala}}]{2018SPIE10700E..0WD}
{Doi}, M., {Miyata}, T., {Yoshii}, Y., {et~al.} 2018, in Society of
  Photo-Optical Instrumentation Engineers (SPIE) Conference Series, Vol. 10700,
  \procspie, 107000W

\bibitem[{{Dong} {et~al.}(2011){Dong}, {Wang}, {Ho}, {Wang}, {Fan}, {Wang},
  {Zhou}, \& {Yuan}}]{2011ApJ...736...86D}
{Dong}, X.-B., {Wang}, J.-G., {Ho}, L.~C., {et~al.} 2011, \apj, 736, 86

\bibitem[{{Dong} {et~al.}(2009){Dong}, {Wang}, {Wang}, {Fan}, {Wang}, {Zhou},
  \& {Yuan}}]{2009ApJ...703L...1D}
{Dong}, X.-B., {Wang}, T.-G., {Wang}, J.-G., {et~al.} 2009, \apjl, 703, L1

\bibitem[{{Feltzing} {et~al.}(2003){Feltzing}, {Bensby}, \&
  {Lundstr{\"o}m}}]{2003A&A...397L...1F}
{Feltzing}, S., {Bensby}, T., \& {Lundstr{\"o}m}, I. 2003, \aap, 397, L1

\bibitem[{{Freudling} {et~al.}(2003){Freudling}, {Corbin}, \&
  {Korista}}]{2003ApJ...587L..67F}
{Freudling}, W., {Corbin}, M.~R., \& {Korista}, K.~T. 2003, \apjl, 587, L67

\bibitem[{{Grandi}(1982)}]{1982ApJ...255...25G}
{Grandi}, S.~A. 1982, \apj, 255, 25

\bibitem[{{Hamann} \& {Ferland}(1999)}]{1999ARA&A..37..487H}
{Hamann}, F., \& {Ferland}, G. 1999, \araa, 37, 487

\bibitem[{{Hamann} {et~al.}(2002){Hamann}, {Korista}, {Ferland}, {Warner}, \&
  {Baldwin}}]{2002ApJ...564..592H}
{Hamann}, F., {Korista}, K.~T., {Ferland}, G.~J., {Warner}, C., \& {Baldwin},
  J. 2002, \apj, 564, 592

\bibitem[{{Ikeda} {et~al.}(2016){Ikeda}, {Kobayashi}, {Kondo}, {Otsubo},
  {Hamano}, {Sameshima}, {Yoshikawa}, {Fukue}, {Nakanishi}, {Kawanishi},
  {Nakaoka}, {Kinoshita}, {Kitano}, {Asano}, {Takenaka}, {Watase}, {Mito},
  {Yasui}, {Minami}, {Izumu}, {Yamamoto}, {Mizumoto}, {Arasaki}, {Arai},
  {Matsunaga}, \& {Kawakita}}]{2016SPIE.9908E..5ZI}
{Ikeda}, Y., {Kobayashi}, N., {Kondo}, S., {et~al.} 2016, in \procspie, Vol.
  9908, Ground-based and Airborne Instrumentation for Astronomy VI, 99085Z

\bibitem[{{Iwamuro} {et~al.}(2004){Iwamuro}, {Kimura}, {Eto}, {Maihara},
  {Motohara}, {Yoshii}, \& {Doi}}]{2004ApJ...614...69I}
{Iwamuro}, F., {Kimura}, M., {Eto}, S., {et~al.} 2004, \apj, 614, 69

\bibitem[{{Iwamuro} {et~al.}(2002){Iwamuro}, {Motohara}, {Maihara}, {Kimura},
  {Yoshii}, \& {Doi}}]{2002ApJ...565...63I}
{Iwamuro}, F., {Motohara}, K., {Maihara}, T., {et~al.} 2002, \apj, 565, 63

\bibitem[{{Jiang} {et~al.}(2008){Jiang}, {Fan}, \&
  {Vestergaard}}]{2008ApJ...679..962J}
{Jiang}, L., {Fan}, X., \& {Vestergaard}, M. 2008, \apj, 679, 962

\bibitem[{{Jiang} {et~al.}(2007){Jiang}, {Fan}, {Vestergaard}, {Kurk},
  {Walter}, {Kelly}, \& {Strauss}}]{2007AJ....134.1150J}
{Jiang}, L., {Fan}, X., {Vestergaard}, M., {et~al.} 2007, \aj, 134, 1150

\bibitem[{{Kausch} {et~al.}(2015){Kausch}, {Noll}, {Smette}, {Kimeswenger},
  {Barden}, {Szyszka}, {Jones}, {Sana}, {Horst}, \&
  {Kerber}}]{2015A&A...576A..78K}
{Kausch}, W., {Noll}, S., {Smette}, A., {et~al.} 2015, \aap, 576, A78

\bibitem[{{Kawara} {et~al.}(1996){Kawara}, {Murayama}, {Taniguchi}, \&
  {Arimoto}}]{1996ApJ...470L..85K}
{Kawara}, K., {Murayama}, T., {Taniguchi}, Y., \& {Arimoto}, N. 1996, \apjl,
  470, L85

\bibitem[{{Korista} {et~al.}(1998){Korista}, {Baldwin}, \&
  {Ferland}}]{1998ApJ...507...24K}
{Korista}, K., {Baldwin}, J., \& {Ferland}, G. 1998, \apj, 507, 24

\bibitem[{{Kurk} {et~al.}(2007){Kurk}, {Walter}, {Fan}, {Jiang}, {Riechers},
  {Rix}, {Pentericci}, {Strauss}, {Carilli}, \& {Wagner}}]{2007ApJ...669...32K}
{Kurk}, J.~D., {Walter}, F., {Fan}, X., {et~al.} 2007, \apj, 669, 32

\bibitem[{{Laor} {et~al.}(1995){Laor}, {Bahcall}, {Jannuzi}, {Schneider}, \&
  {Green}}]{1995ApJS...99....1L}
{Laor}, A., {Bahcall}, J.~N., {Jannuzi}, B.~T., {Schneider}, D.~P., \& {Green},
  R.~F. 1995, \apjs, 99, 1

\bibitem[{{Maiolino} {et~al.}(2003){Maiolino}, {Juarez}, {Mujica}, {Nagar}, \&
  {Oliva}}]{2003ApJ...596L.155M}
{Maiolino}, R., {Juarez}, Y., {Mujica}, R., {Nagar}, N.~M., \& {Oliva}, E.
  2003, \apjl, 596, L155

\bibitem[{{Markwardt}(2009)}]{2009ASPC..411..251M}
{Markwardt}, C.~B. 2009, in Astronomical Society of the Pacific Conference
  Series, Vol. 411, Astronomical Data Analysis Software and Systems XVIII, ed.
  D.~A. {Bohlender}, D.~{Durand}, \& P.~{Dowler}, 251

\bibitem[{{Mazzucchelli} {et~al.}(2017){Mazzucchelli}, {Ba{\~n}ados},
  {Venemans}, {Decarli}, {Farina}, {Walter}, {Eilers}, {Rix}, {Simcoe},
  {Stern}, {Fan}, {Schlafly}, {De Rosa}, {Hennawi}, {Chambers}, {Greiner},
  {Burgett}, {Draper}, {Kaiser}, {Kudritzki}, {Magnier}, {Metcalfe}, {Waters},
  \& {Wainscoat}}]{2017ApJ...849...91M}
{Mazzucchelli}, C., {Ba{\~n}ados}, E., {Venemans}, B.~P., {et~al.} 2017, \apj,
  849, 91

\bibitem[{{McWilliam}(1997)}]{1997ARA&A..35..503M}
{McWilliam}, A. 1997, \araa, 35, 503

\bibitem[{{Mushotzky} \& {Ferland}(1984)}]{1984ApJ...278..558M}
{Mushotzky}, R., \& {Ferland}, G.~J. 1984, \apj, 278, 558

\bibitem[{{Netzer}(1985)}]{1985MNRAS.216...63N}
{Netzer}, H. 1985, \mnras, 216, 63

\bibitem[{{Onoue} {et~al.}(2020){Onoue}, {Ba{\~n}ados}, {Mazzucchelli},
  {Venemans}, {Schindler}, {Walter}, {Hennawi}, {Andika}, {Davies}, {Decarli},
  {Farina}, {Jahnke}, {Nagao}, {Tominaga}, \& {Wang}}]{2020ApJ...898..105O}
{Onoue}, M., {Ba{\~n}ados}, E., {Mazzucchelli}, C., {et~al.} 2020, \apj, 898,
  105

\bibitem[{{Osmer} \& {Shields}(1999)}]{1999ASPC..162..235O}
{Osmer}, P.~S., \& {Shields}, J.~C. 1999, in Astronomical Society of the
  Pacific Conference Series, Vol. 162, Quasars and Cosmology, ed. G.~{Ferland}
  \& J.~{Baldwin}, 235

\bibitem[{{P{\^a}ris} {et~al.}(2017){P{\^a}ris}, {Petitjean}, {Ross}, {Myers},
  {Aubourg}, {Streblyanska}, {Bailey}, {Armengaud}, {Palanque-Delabrouille},
  {Y{\`e}che}, {Hamann}, {Strauss}, {Albareti}, {Bovy}, {Bizyaev}, {Niel
  Brandt}, {Brusa}, {Buchner}, {Comparat}, {Croft}, {Dwelly}, {Fan},
  {Font-Ribera}, {Ge}, {Georgakakis}, {Hall}, {Jiang}, {Kinemuchi},
  {Malanushenko}, {Malanushenko}, {McMahon}, {Menzel}, {Merloni}, {Nandra},
  {Noterdaeme}, {Oravetz}, {Pan}, {Pieri}, {Prada}, {Salvato}, {Schlegel},
  {Schneider}, {Simmons}, {Viel}, {Weinberg}, \& {Zhu}}]{2017AA...597A..79P}
{P{\^a}ris}, I., {Petitjean}, P., {Ross}, N.~P., {et~al.} 2017, \aap, 597, A79

\bibitem[{{Peterson}(1997)}]{1997iagn.book.....P}
{Peterson}, B.~M. 1997, {An Introduction to Active Galactic Nuclei} (New York
  Cambridge University Press)

\bibitem[{{Pillepich} {et~al.}(2018){Pillepich}, {Nelson}, {Hernquist},
  {Springel}, {Pakmor}, {Torrey}, {Weinberger}, {Genel}, {Naiman}, {Marinacci},
  \& {Vogelsberger}}]{2018MNRAS.475..648P}
{Pillepich}, A., {Nelson}, D., {Hernquist}, L., {et~al.} 2018, \mnras, 475, 648

\bibitem[{{Rayner} {et~al.}(2009){Rayner}, {Cushing}, \&
  {Vacca}}]{2009ApJS..185..289R}
{Rayner}, J.~T., {Cushing}, M.~C., \& {Vacca}, W.~D. 2009, \apjs, 185, 289

\bibitem[{{Ross} {et~al.}(2013){Ross}, {McGreer}, {White}, {Richards}, {Myers},
  {Palanque-Delabrouille}, {Strauss}, {Anderson}, {Shen}, {Brandt},
  {Y{\`e}che}, {Swanson}, {Aubourg}, {Bailey}, {Bizyaev}, {Bovy}, {Brewington},
  {Brinkmann}, {DeGraf}, {Di Matteo}, {Ebelke}, {Fan}, {Ge}, {Malanushenko},
  {Malanushenko}, {Mandelbaum}, {Maraston}, {Muna}, {Oravetz}, {Pan},
  {P{\^a}ris}, {Petitjean}, {Schawinski}, {Schlegel}, {Schneider}, {Silverman},
  {Simmons}, {Snedden}, {Streblyanska}, {Suzuki}, {Weinberg}, \&
  {York}}]{2013ApJ...773...14R}
{Ross}, N.~P., {McGreer}, I.~D., {White}, M., {et~al.} 2013, \apj, 773, 14

\bibitem[{{Sameshima} {et~al.}(2017){Sameshima}, {Yoshii}, \&
  {Kawara}}]{2017ApJ...834..203S}
{Sameshima}, H., {Yoshii}, Y., \& {Kawara}, K. 2017, \apj, 834, 203 (Paper~I)

\bibitem[{{Sameshima} {et~al.}(2009){Sameshima}, {Maza}, {Matsuoka}, {Oyabu},
  {Kawara}, {Yoshii}, {Asami}, {Ienaka}, \& {Tsuzuki}}]{2009MNRAS.395.1087S}
{Sameshima}, H., {Maza}, J., {Matsuoka}, Y., {et~al.} 2009, \mnras, 395, 1087

\bibitem[{{Schaye} {et~al.}(2015){Schaye}, {Crain}, {Bower}, {Furlong},
  {Schaller}, {Theuns}, {Dalla Vecchia}, {Frenk}, {McCarthy}, {Helly},
  {Jenkins}, {Rosas-Guevara}, {White}, {Baes}, {Booth}, {Camps}, {Navarro},
  {Qu}, {Rahmati}, {Sawala}, {Thomas}, \& {Trayford}}]{2015MNRAS.446..521S}
{Schaye}, J., {Crain}, R.~A., {Bower}, R.~G., {et~al.} 2015, \mnras, 446, 521

\bibitem[{{Schneider} {et~al.}(2010){Schneider}, {Richards}, \&
  {Hall}}]{2010AJ....139.2360S}
{Schneider}, D.~P., {Richards}, G.~T., \& {Hall}, P.~B. 2010, \aj, 139, 2360

\bibitem[{{Shang} {et~al.}(2003){Shang}, {Wills}, {Robinson}, {Wills}, {Laor},
  {Xie}, \& {Yuan}}]{2003ApJ...586...52S}
{Shang}, Z., {Wills}, B.~J., {Robinson}, E.~L., {et~al.} 2003, \apj, 586, 52

\bibitem[{{Shen} {et~al.}(2011){Shen}, {Richards}, {Strauss}, {Hall},
  {Schneider}, {Snedden}, {Bizyaev}, {Brewington}, {Malanushenko},
  {Malanushenko}, {Oravetz}, {Pan}, \& {Simmons}}]{2011ApJS..194...45S}
{Shen}, Y., {Richards}, G.~T., {Strauss}, M.~A., {et~al.} 2011, \apjs, 194, 45

\bibitem[{{Shields}(2007)}]{2007ASPC..373..355S}
{Shields}, J.~C. 2007, in Astronomical Society of the Pacific Conference
  Series, Vol. 373, The Central Engine of Active Galactic Nuclei, ed. L.~C.
  {Ho} \& J.~W. {Wang}, 355

\bibitem[{{Shin} {et~al.}(2019){Shin}, {Nagao}, {Woo}, \&
  {Le}}]{2019ApJ...874...22S}
{Shin}, J., {Nagao}, T., {Woo}, J.-H., \& {Le}, H. A.~N. 2019, \apj, 874, 22

\bibitem[{{Smette} {et~al.}(2015){Smette}, {Sana}, {Noll}, {Horst}, {Kausch},
  {Kimeswenger}, {Barden}, {Szyszka}, {Jones}, {Gallenne}, {Vinther},
  {Ballester}, \& {Taylor}}]{2015A&A...576A..77S}
{Smette}, A., {Sana}, H., {Noll}, S., {et~al.} 2015, \aap, 576, A77

\bibitem[{{Sulentic} {et~al.}(2000){Sulentic}, {Zwitter}, {Marziani}, \&
  {Dultzin-Hacyan}}]{2000ApJ...536L...5S}
{Sulentic}, J.~W., {Zwitter}, T., {Marziani}, P., \& {Dultzin-Hacyan}, D. 2000,
  \apjl, 536, L5

\bibitem[{{Thompson} {et~al.}(1999){Thompson}, {Hill}, \&
  {Elston}}]{1999ApJ...515..487T}
{Thompson}, K.~L., {Hill}, G.~J., \& {Elston}, R. 1999, \apj, 515, 487

\bibitem[{{Tinsley}(1979)}]{1979ApJ...229.1046T}
{Tinsley}, B.~M. 1979, \apj, 229, 1046

\bibitem[{{Tolstoy} {et~al.}(2009){Tolstoy}, {Hill}, \&
  {Tosi}}]{2009ARA&A..47..371T}
{Tolstoy}, E., {Hill}, V., \& {Tosi}, M. 2009, \araa, 47, 371

\bibitem[{{Tsuzuki} {et~al.}(2006){Tsuzuki}, {Kawara}, {Yoshii}, {Oyabu},
  {Tanab{\'e}}, \& {Matsuoka}}]{2006ApJ...650...57T}
{Tsuzuki}, Y., {Kawara}, K., {Yoshii}, Y., {et~al.} 2006, \apj, 650, 57

\bibitem[{{Vestergaard} \& {Osmer}(2009)}]{2009ApJ...699..800V}
{Vestergaard}, M., \& {Osmer}, P.~S. 2009, \apj, 699, 800

\bibitem[{{Vogelsberger} {et~al.}(2014){Vogelsberger}, {Zavala}, {Simpson}, \&
  {Jenkins}}]{2014MNRAS.444.3684V}
{Vogelsberger}, M., {Zavala}, J., {Simpson}, C., \& {Jenkins}, A. 2014, \mnras,
  444, 3684

\bibitem[{{Warner} {et~al.}(2003){Warner}, {Hamann}, \&
  {Dietrich}}]{2003ApJ...596...72W}
{Warner}, C., {Hamann}, F., \& {Dietrich}, M. 2003, \apj, 596, 72

\bibitem[{{Warner} {et~al.}(2004){Warner}, {Hamann}, \&
  {Dietrich}}]{2004ApJ...608..136W}
---. 2004, \apj, 608, 136

\bibitem[{{Wills} {et~al.}(1985){Wills}, {Netzer}, \&
  {Wills}}]{1985ApJ...288...94W}
{Wills}, B.~J., {Netzer}, H., \& {Wills}, D. 1985, \apj, 288, 94

\bibitem[{{Xu} {et~al.}(2008){Xu}, {Bian}, {Yuan}, \&
  {Huang}}]{2008MNRAS.389.1703X}
{Xu}, Y., {Bian}, W.-H., {Yuan}, Q.-R., \& {Huang}, K.-L. 2008, \mnras, 389,
  1703

\bibitem[{{Yoshii} {et~al.}(1998){Yoshii}, {Tsujimoto}, \&
  {Kawara}}]{1998ApJ...507L.113Y}
{Yoshii}, Y., {Tsujimoto}, T., \& {Kawara}, K. 1998, \apjl, 507, L113

\bibitem[{{Yoshii} {et~al.}(1996){Yoshii}, {Tsujimoto}, \&
  {Nomoto}}]{1996ApJ...462..266Y}
{Yoshii}, Y., {Tsujimoto}, T., \& {Nomoto}, K. 1996, \apj, 462, 266

\bibitem[{{Yoshii} {et~al.}(2002){Yoshii}, {Doi}, {Handa}, {Kawara}, {Kohno},
  {Minezaki}, {Mitsuda}, {Miyata}, {Motohara}, \&
  {Tanaka}}]{2002aprm.conf...35Y}
{Yoshii}, Y., {Doi}, M., {Handa}, T., {et~al.} 2002, in 8th Asian-Pacific
  Regional Meeting, Volume II, ed. S.~{Ikeuchi}, J.~{Hearnshaw}, \&
  T.~{Hanawa}, 35--36

\bibitem[{{Yoshii} {et~al.}(2014){Yoshii}, {Doi}, {Kohno}, {Miyata},
  {Motohara}, {Kawara}, {Tanaka}, {Minezaki}, {Sako}, {Morokuma}, {Tamura},
  {Tanabe}, {Takahashi}, {Konishi}, {Kamizuka}, {Koshida}, {Kato}, {Aoki},
  {Soyano}, {Tarusawa}, {Hand a}, {Bronfman}, {Ruiz}, {Hamuy}, \&
  {Mendez}}]{2014SPIE.9145E..07Y}
{Yoshii}, Y., {Doi}, M., {Kohno}, K., {et~al.} 2014, in Society of
  Photo-Optical Instrumentation Engineers (SPIE) Conference Series, Vol. 9145,
  \procspie, 914507

\bibitem[{{Yoshii} {et~al.}(2016){Yoshii}, {Doi}, {Kohno}, {Miyata},
  {Motohara}, {Kawara}, {Tanaka}, {Minezaki}, {Sako}, {Morokuma}, {Tamura},
  {Tanabe}, {Takahashi}, {Konishi}, {Kamizuka}, {Kato}, {Aoki}, {Soyano},
  {Tarusawa}, {Handa}, {Koshida}, {Bronfman}, {Ruiz}, {Hamuy}, \&
  {Garay}}]{2016SPIE.9906E..0RY}
{Yoshii}, Y., {Doi}, M., {Kohno}, K., {et~al.} 2016, in Society of
  Photo-Optical Instrumentation Engineers (SPIE) Conference Series, Vol. 9906,
  \procspie, 99060R

\end{thebibliography}

\bibliographystyle{apj}



\end{document}